\catcode`\@=11
\newif\if@fewtab\@fewtabtrue

{\count255=\time\divide\count255 by 60
\xdef\hourmin{\number\count255}
\multiply\count255 by-60\advance\count255 by\time
\xdef\hourmin{\hourmin:\ifnum\count255<10 0\fi\the\count255}}
\def\ps@draft{\let\@mkboth\@gobbletwo
    \def\@oddhead{}
    \def\@oddfoot
       {\hbox to 7 cm{$\scriptstyle Draft\ version:\ \draftdate$
       \hfil}\hskip -7cm\hfil\rm\thepage \hfil}
    \def\@evenhead{}\let\@evenfoot\@oddfoot}

\def\ceqno{\global\@fewtabfalse
    \ifcase\@eqcnt \def\@tempa{& & &}\or \def\@tempa{& &}
      \or \def\@tempa{&}
      \or\def\@tempa{}\fi\@tempa
{\rm(\theequation)}}

\def\aeqno#1{\global\@fewtabfalse
    \ifcase\@eqcnt \def\@tempa{& & &}\or \def\@tempa{& &}
      \or \def\@tempa{&}
      \or\def\@tempa{}\fi\@tempa
{\rm(\theequation,#1)}}

\def\label#1{\ifnum\draftcontrol=1
 \global\def\draftnote{$\scriptstyle #1$}\fi
 \@bsphack\if@filesw {\let\thepage\relax
   \def\protect{\noexpand\noexpand\noexpand}%
\xdef\@gtempa{\write\@auxout{\string
      \newlabel{#1}{{\@currentlabel}{\thepage}}}}}\@gtempa
   \if@nobreak \ifvmode\nobreak\fi\fi\fi
  \@esphack}

\def\alabel#1#2{\label{#1}\global\@fewtabfalse
    \ifcase\@eqcnt \def\@tempa{& & &}\or \def\@tempa{& &}
      \or \def\@tempa{&}
      \or\def\@tempa{}\fi\@tempa
{\hbox to 3cm{\phantom{\rm(\theequation,#2)}
\draftnote \hfil}\hskip -3cm {\rm(\theequation,#2)}}}

\def\clabel#1{\label{#1}\global\@fewtabfalse
    \ifcase\@eqcnt \def\@tempa{& & &}\or \def\@tempa{& &}
      \or \def\@tempa{&}
      \or\def\@tempa{}\fi\@tempa
{\hbox to 3cm{\phantom{\rm(\theequation)}
\draftnote \hfil}\hskip -3cm{\rm(\theequation)}}}

\def\eqnarray{\def\draftnote{{}}\global\@fewtabtrue
\stepcounter{equation}\let\@currentlabel=\theequation
\global\@eqnswtrue
\global\@eqcnt\z@\tabskip\@centering\let\\=\@eqncr
$$\halign to \displaywidth\bgroup\@eqnsel\hskip\@centering\@eqcnt\z@
  $\displaystyle\tabskip\z@{##}$&\global\@eqcnt\@ne
  \hskip 1\arraycolsep \hfil${##}$\hfil
  &\global\@eqcnt\tw@ \hskip 1\arraycolsep
$\displaystyle\tabskip\z@{##}$
\hfil  \tabskip\@centering&\global\@eqcnt\thr@@\llap{##}\tabskip\z@
\cr}

\def\endeqnarray{\@@eqncr\egroup
      \global\advance\c@equation\m@ne$$\global\@ignoretrue}

\def\@eqnnum{\hbox to 3cm{\phantom{\rm(\theequation)} \draftnote
                         \hfil}\hskip -3cm {\rm(\theequation)}}

\def\@@eqncr{\let\@tempa\relax
    \ifcase\@eqcnt \def\@tempa{& & &}\or \def\@tempa{& &}
      \or \def\@tempa{&}
      \or\def\@tempa{}
\fi\@tempa
\if@eqnsw
\if@fewtab\@eqnnum\fi
\stepcounter{equation}\fi\global
\@eqnswtrue\global\@eqcnt\z@\global\@fewtabtrue\cr}

\def\draftcite#1{\ifnum\draftcontrol=1#1\else{}\fi}

\def\@lbibitem[#1]#2{\item{}\hskip -3cm \hbox to 2cm
{\hfil$\scriptstyle\draftcite{#2}$}\hskip
1cm[\@biblabel{#1}]\if@filesw
     {\def\protect##1{\string ##1\space}\immediate
      \write\@auxout{\string\bibcite{#2}{#1}}}\fi\ignorespaces}

\def\@bibitem#1{\item\hskip -3cm \hbox to 2cm
{\hfil $\scriptstyle\draftcite{#1}$}\hskip 1cm
\if@filesw \immediate\write\@auxout
       {\string\bibcite{#1}{\the\value{\@listctr}}}\fi\ignorespaces}

\def\nsection#1{\section{#1}\setcounter{equation}{0}}

\def\nappendix#1{\vskip 0.8cm\no{\bf Appendix #1}\def\thesection{#1}
\setcounter{equation}{0}}

\font\tendl=msbm10  scaled \magstep1
\font\sevendl=msbm7 scaled \magstep1
\font\fivedl=msbm5 scaled \magstep1
\font\tengl=eufm10  scaled \magstep1
\font\sevengl=eufm7 scaled \magstep1
\font\fivegl=eufm5 scaled \magstep1

\newfam\dlfam  
\textfont\dlfam=\tendl \scriptfont\dlfam=\sevendl
\scriptscriptfont\dlfam=\fivedl
\newfam\glfam  
\textfont\glfam=\tengl \scriptfont\glfam=\sevengl
\scriptscriptfont\glfam=\fivegl

\def\draftdate{\number\month/\number\day/\number\year\ \ \ \hourmin }

\global\def\draftcontrol{0}
\catcode`\@=12
\def\tilde{\widetilde}

\documentstyle[12pt,aps,epsfig]{revtex}
\newcommand{\bm}{\bbox}
\newcommand{\qq}{\begin{eqnarray}}
\newcommand{\qqq}{\end{eqnarray}}

\renewcommand{\thesection}{\arabic{section}}

\def\theequation{{\thesection.\arabic{equation}}}

\renewcommand{\baselinestretch}{1.0}
\newcommand{\be}{\begin{eqnarray}}
\newcommand{\en}{\end{eqnarray}\vs 0.5 cm}

\newcommand{\no}{\noindent}
\newcommand{\vs}{\vskip}

\newcommand{\ee}{{\rm e}}

\newcommand{\CD}{{\cal D}}

\newcommand{\CH}{{\cal H}}

\newcommand{\CK}{{\cal K}}
\newcommand{\CL}{{\cal L}}

\newcommand{\CN}{{\cal N}}
\newcommand{\CO}{{\cal O}}
\newcommand{\CP}{{\cal P}}
\newcommand{\CQ}{{\cal Q}}

\newcommand{\CW}{{\cal W}}

\newcommand{\m}{\hspace{0.025cm}}

\newcommand{\br}{{\bm r}}
\newcommand{\brho}{{\bm\rho}}
\newcommand{\bR}{{\bm R}}
\newcommand{\bv}{{\bm v}}

\begin{document}
\title{\Large{\bf{Lagrangian dispersion in Gaussian\\ self-similar
velocity ensembles\\{\ }}}}
\author{Marta Chaves$^{1}$,\ \,\,Krzysztof Gaw\c{e}dzki$^{*,2}$,\ \,
Peter Horvai$^{2,3}$,\\ Antti Kupiainen$^{4}$, \ \,\,Massimo
Vergassola$^{*,1}$
\\ }
\address{{\ }\\ $^{*}$\,member of CNRS \\ $^{1}$Observatoire de la C\^{o}te
d'Azur, B.P. 4229, 06304 Nice,
France\\ $^{2}$Laboratoire de Physique, ENS-Lyon, 46 All\'ee d'Italie,
69364 Lyon, France\\ $^{3}$Centre de Physique Th\'eorique,
Ecole Polytechnique, 91128 Palaisseau, France\\
$^{4}$Department of Mathematics, Helsinki University, P.O. Box 4,
00014 Helsinki,\\ Finland}

\vskip -0.5cm

\maketitle

\vskip 1.3cm

\begin{center}
{\bf Abstract}
\end{center}

\begin{abstract}
We analyze the Lagrangian flow in a family of simple
Gaussian scale-invariant velocity ensembles that exhibit both
spatial roughness and temporal correlations. We show that the
behavior of the Lagrangian dispersion of pairs of fluid particles
in such models is determined by the scale dependence of the ratio
between the correlation time of velocity differences and the eddy
turnover time. For a non-trivial scale dependence, the asymptotic
regimes of the dispersion at small and large scales are described
by the models with either rapidly decorrelating or frozen velocities.
In contrast to the decorrelated case, known as the Kraichnan model
and exhibiting Lagrangian flows with deterministic or stochastic 
trajectories, fast separating or trapped together, the frozen 
model is poorly understood. We examine the pair dispersion behavior 
in its simplest, one-dimensional version, reinforcing analytic 
arguments by numerical analysis. The collected information about 
the pair dispersion statistics in the limiting models allows to 
partially predict the extent of different phases and the scaling 
properties of the Lagrangian flow in the model with time-correlated 
velocities.
\end{abstract}
\vskip 1.4cm

\nsection{Introduction} \label{sec:INTRO}

\noindent The aim of this paper is to study the Lagrangian flow 
in $d$-dimensional random velocity fields $\,\bv(t,\br)\,$
with a prescribed scale-invariant statistics. The velocity 
ensembles that we shall consider mimic some essential properties 
of realistic velocities in developed turbulence:  their 
spatial roughness within a large interval of scales and 
their temporal correlation. By definition, the Lagrangian 
flow is described by the ordinary differential equation \qq
{d\bR\over dt}\ =\ \bv(t,\bR)\,. \label{Lf} \qqq It determines the
motion of hypothetical fluid particles or of small test particles
suspended in the fluid. One usually distinguishes between the
motion of a single particle, dominated by the velocity
fluctuations on the largest scale present (the so called "sweeping
effects") and the evolution of a relative separation of two
particles, the object of interest of the present paper, driven 
by the velocity fluctuations on scales of the order of the 
inter-particle distance. For larger groups of particles, one should 
similarly distinguish the motion of their barycenter from 
the relative motion of particles within the group. The latter is 
known to show quite intricate behavior related to intermittency, 
see \cite{RMP}, but it will not interest us here.  
\vskip 0.5cm

The separation $\,\brho=\bR'-\bR\,$ between two fluid particles
satisfies the equation \qq{d\brho\over
dt}=\bv(t,\brho+\bR(t))-\bv(t,\bR(t))
\,,\label{rho}\qqq where $\bR(t)$ is a trajectory of one of the
particles, a solution of Eq.\,\,(\ref{Lf}) starting at time zero
at $\bR=0$, \,for example. Upon introduction of the so called
quasi-Lagrangian velocity, \qq
\bv^{qL}(t,\br)=\bv(t,\br+\bR(t))\,,\label{qL}\qqq
i.e.\,\,velocity in the frame moving with a fixed fluid particle,
we may rewrite Eq.\,\,(\ref{rho}) as \qq{d\brho\over
dt}=\bv^{qL}(t,\brho)-\bv^{qL}(t,\bm{0})\equiv\Delta
\bv^{qL}(t,\brho)
\,.\label{rho1}\qqq
We shall be interested in the short- and long-time behaviors of
the Lagrangian particles in the statistical ensembles where
typical velocities are only H\"older continuous in space. In such
non-Lipschitz velocities, there is a problem with solving Eqs.
(\ref{Lf}), (\ref{rho}) or (\ref{rho1}). To avoid it, we shall
first consider noisy particle trajectories that solve the
stochastic equation \qq d{\bm R}\ =\ {\bm v} (t,{\bm
R})\,dt\,+\,\sqrt{2\kappa}\,d{\bm W}\,, \label{nse} \qqq where
$\,{\bm W}(t)\,$ is the Brownian motion in $\,d\,$ dimensions. Noisy
trajectories form a well defined Markov process even in velocity
fields with poor regularity. Subsequently, the limit $\kappa\to0$
will be performed in selected quantities. The velocity ensembles
that we shall consider are time-reversal invariant. As a result,
we shall not have to distinguish the forward and the backward
evolution of trajectories and will concentrate on the first one,
between, say, times $0$ and $t$.
\vskip 0.5cm

One way to study the relative motion of pairs of fluid
particles is to follow the evolution of the {\bf pair dispersion},
i.e.\,\,of the separation distance $\,\rho\,$ between two
particles. Its statistics
in a random flow may be described by the velocity-averaged
probability distribution $\,\CP(\rho_0,\rho;t)\,d\rho\,$ of the
time $t$ dispersion $\,\rho\,$ given its time zero value
$\,\rho_0$, \,in the limit when we remove the (independent) noises
of the Lagrangian trajectories. As we shall see, the PDF's 
$\,\CP(\rho_0,\rho;t)\,$ may be sometimes distributions rather than integrable
functions.
Another, related, test of the relative motion of a pair of fluid
particles is obtained by looking  at the {\bf exit time}
\cite{GZ,BCCV}: the time $\,t\,$ that the pair dispersion takes to
evolve from $\,\rho_0$ to $\,\rho_1$. In particular,
$\,\rho_1=2\rho_0\,$ corresponds to the {\bf doubling time} of 
the pair dispersion. The statistics of the exit times may be encoded 
in their velocity-averaged probability distribution $\,\CQ(t;\rho_0,
\rho_1)\,dt\,$ taken in the limit of vanishing noise. The exit
time is less influenced than the pair dispersion by small or
large-distance cutoffs in the velocity correlations, so preferable
in numerical or experimental studies \cite{BCCV}. \vskip 0.5cm

Recently, a new insight into the intricate character of the
Lagrangian flow in turbulent velocities has been gained, see
\cite{SM,GV,LeJR1,EVDE1,LeJR2}, by analytic study of 
the Kraichnan ensemble \cite{Kr68} of Gaussian velocities
which are decorrelated in time but exhibit scaling behavior 
in space. Here, we try to find out how the presence 
of temporal correlations of velocities influences the Lagrangian 
flow. We shall study the behavior of trajectory separation in a simple
generalization of the Kraichnan ensemble of velocities where time
correlations are reintroduced. \vskip 0.5cm

The paper is organized as follows. In Sect.\,\ref{sec:LKR} we
recall the main facts about the Lagrangian flow in the Kraichnan
model, in particular the appearance of phases with very different
trajectory behavior. 
Sect.\,\ref{sec:GVTC} describes a Gaussian ensemble of velocities
with temporal correlations, discussed in the past in
\cite{CFL,Ant,F1,FK}. It exposes a simple mean-field type
analysis of the particle separation when such an ensemble is used to
model the quasi-Lagrangian velocities. How the mean-field predictions
may be substantiated further by scaling arguments is the subject of
Sect.\,\ref{sec:SA}. See also \cite{F2,F3} for related rigorous 
results. Analytic arguments and conjectures about the behavior 
of trajectories in a one-dimensional version of the model 
with time-independent velocities are contained in Sect.\,\ref{sec:FR1D}.
The behavior of the exit time statistics in velocity ensembles with
long-time correlations is studied in Sect.\,\,\ref{sec:LTS}. 
The question how the behavior of pair dispersion changes when the Gaussian
ensemble is used to model the Eulerian velocities is addressed in
Sect.\,\ref{sec:EUL}. In particular we show that in the
one-dimensional time-independent case, the sweeping by large
eddies in the Eulerian model speeds up the movement of a single
Lagrangian particle but it localizes pairs of particles by reducing
the growth of their dispersion. 
Three Appendices contain more technical
material. Some of the predictions of the paper are checked in one dimension 
by numerical simulations.
\vskip 0.6cm

{\bf Acknowledgements}: The work of MC was supported by the  
Funda\c{c}\~ao para a Ci\^encia e a Tecnologia grant: PRAXIS XXI/BD/21413/99.
KG, PH, AK and MV are grateful to the Erwin
Schr\"{o}dinger Institute in Vienna and the Institute for Advanced Study in Princeton
where parts of the work on the paper were done. KG acknowledges the support of 
the Neumann Fund and MV the Ralph E. and Doris M. Hansmann Membership at the IAS.
AK and MV were also supported by the European Union under the Contracts FMRX-CT98-0175 and
HPRN-CT-2000-00162, respectively.
\vskip 0.5cm

\nsection{Lessons from the Kraichnan model} \label{sec:LKR}

\noindent The {\bf Kraichnan ensemble} of turbulent velocities \cite{Kr68},
is a Gaussian ensemble with vanishing velocity 1-point function
and with the 2-point function \qq \Big\langle v^i(t,{\bm
r})\,v^j(t',{\bm r}')\Big\rangle \ =\ D_1\,\,\delta(t-t')\int{{\rm
e}^{\,\,i\,{\bm k \cdot({\bm r}-{\bm r}')}}
\over{k_{_L}^{d+\xi}}}\,\,P^{ij}({\bm k},\wp) \,\,{{d{\bm
k}}\over{(2\pi)^d}}\,, \label{vcKr} \qqq where
$\,k_{_L}=\sqrt{k^2+L^{-2}}\,$ and $\,P^{ij}({\bm k},\wp) ={1-\wp
\over d-1}\Big(\delta^{ij}-{k^ik^j\over k^2}\Big)+\wp{k^ik^j\over k^2}$, see
\cite{RMP} and references therein. There are two dimensionless
parameters in the Kraichnan ensemble: the exponent $\,\xi>0\,$ and
the compressibility degree $\,0\leq\wp\leq 1$. For $\,\xi\leq2$,
the velocities smeared in time are (almost surely) H\"older
continuous in space with any exponent smaller than $\xi/2$. For
$\,\xi>2$, they are Lipschitz (or even more regular). The
compressibility degree $\,\wp=0\,$ corresponds to incompressible
velocities, $\,\wp=1\,$ to gradients of a potential (in one
dimension necessarily $\wp=1$). \,The normalization constant $\,D_1\,$
has dimension $\,length^{2-\xi}/time$. \,The length $L$ is the
integral scale that sets the spatial correlation length of
velocities. If $\,\xi<2$, \,it may be taken to infinity in the
correlation functions involving only velocity differences $\,{\bm
v}(t,\br+\brho)-{\bm v}(t,{\bm r})\equiv \Delta{\bm v}(t,\brho)$.
\,For $\,\xi\geq2$, this may still be done if $\,D_1\,$ is rescaled
when $\,L\to\infty$.
\vskip -0.7truecm
\

\subsection{\ Possible flow behaviors}

\noindent The Kraichnan ensemble may be used invariably to model Eulerian or
quasi-Lagrangian velocities as both ensembles coincide in this
case. The statistics of a single Lagrangian particle is that of a
$d$-dimensional Brownian motion with diffusivity that blows up when
$\,L\to0$. \,The two-particle separation depends only on velocity
differences and its statistics has a non-trivial $\,L\to\infty\,$
limit whenever this holds for the velocity differences. The
pair-dispersion and the exit time PDF's $\,\CP(\rho_0,\rho;t)\,$ and
$\,\CQ(t;\rho_0,\rho_1)\,$ may be solved analytically 
in this limit. The exact solutions, that describes also the
short-distance asymptotics of the PDF's at finite $\,L$, \,show
several dichotomic behaviors depending on the values of parameters
of the model. The first dichotomy, noticed in \cite{SM}, is
between the \vskip 0.5cm

\noindent{\bf\ \ \ \ \ deterministic flow}\quad characterized by
the property \qq \lim\limits_{\rho_0\to0}\,\,\CP(\rho_0,\rho;t)\
=\ \delta(\rho)\label{deter} \qqq \noindent which signals that the
trajectories in a fixed velocity field are defined by their
initial position, and the \vskip 0.5cm

\noindent{\bf\ \ \ \ \ stochastic flow}\quad where \qq
\lim\limits_{\rho_0\to0}\,\,\CP(\rho_0,\rho;t)\quad{\rm is\ an\
integrable\ function\ of\ }\ \,\rho\,.  \label{stoch} \qqq
\vskip 0.1cm

\noindent The limits of the PDF's above (and below) should be
understood in weak sense, under integrals against test functions.
The behavior (\ref{stoch}) means that infinitesimally close
trajectories separate in a finite time and indicates that the
stochasticity introduced into the Lagrangian flow by coupling it
to the noise, see Eq.\,\,(\ref{nse}), survives in the limit
$\kappa\to\nobreak0$. \,The Lagrangian trajectories in a fixed velocity
field are not determined by initial position but form instead a
stochastic process. That this is indeed what happens in the Kraichnan
model was established rigorously in \cite{LeJR1}. We shall call
the phenomenon {\bf spontaneous stochasticity}\footnote{It was
termed intrinsic stochasticity in \cite{EVDE1}}. \vskip 0.4cm

There are further dichotomic behaviors of the Lagrangian flow in
the Kraichnan model. We have chosen to characterize the other 
dichotomies in terms of the small $\,\rho_0\,$ behavior of the exit 
time PDF $\,\CQ(t;\rho_0,\gamma\rho_0)\,$ with $\,0<\gamma={\rm
const}.\not=1$, \,attaching to them more or less suggestive names.
\,There is a dichotomy between the \vskip 0.5cm

\noindent{\bf\ \ \ \ \ Lyapunov flow}\quad such that  \qq
\lim\limits_{\rho_0\to0}\,\,\CQ(t;\rho_0,\gamma\rho_0)\quad{\rm is\
an\ integrable\ function\ of}\ \,t\,,\label{lyap} \qqq
\vskip -0.4cm
\noindent and the \vskip 0.5cm

\noindent{\bf\ \ \ \ \ Richardson flow}\quad where \qq
\lim\limits_{\rho_0\to0}\,\,\CQ(t;\rho_0,\gamma\rho_0)\ =\
c(\gamma)\,\delta(t)\quad{\rm with\ }\
c(\gamma)>0\,.\label{Rich}\qqq

\vskip 0.1cm \noindent This dichotomy distinguishes the flows in
regular velocities where the Lagrangian separation on short
distances involves fixed time scales (like the inverse Lyapunov
exponent), from the ones in non-regular (non-Lipschitz) velocities
where the characteristic times of the Lagrangian separation become
very short on short scales. \vskip 0.4cm

Finally, the last two dichotomies that we want to single out
characterize the short distance behavior of the probability of infinite
exit times. They are between the \vskip 0.5cm

\noindent{\bf\ \ \ \ \ locally separating} \ and
\ {\bf locally trapping flow}\quad where for $\,\gamma>1$ \qq
\lim\limits_{\rho_0\to0}\int\CQ(t;\rho_0,\gamma\rho_0)\,dt\ \cases{\
=\ \cr\ <\ }\Big\}\ 1\,,\ \,{\rm respectively},\label{trans}\qqq
\vskip-0.4cm
\noindent and \vskip 0.5cm \noindent{\bf\ \ \ \ \ locally recurrent} \ and
\ {\bf locally transient flow}\quad where the same holds for
$\,\gamma<1$. \vskip 0.5cm

\noindent Roughly, with positive probability, close trajectories
do not increase their distance in locally trapping flows and 
do not approach each other in locally transient ones.

\subsection{\ Pair dispersion}

\noindent Due to the temporal decorrelation of the Kraichnan velocities, 
the probability measures $\,\CP(\rho_0,\rho;t)\,d\rho\,$ constitute
transition probabilities of a Markov process $\,\rho(t)\,$ that in
the limit $\,L\to\infty\,$ becomes a diffusion on a half-line 
with the explicit generator \qq M\ =\
-D_1'\,\rho^{\xi-a}\partial_\rho\,\rho^a
\partial_\rho\qquad{\rm for}\ \quad a\,=\,{_{d+\xi}\over^{1+\wp\xi}}-1\,,
\label{M} \qqq where $\,D_1'$ is proportional to $D_1$. \,Note that
the symbol of $\,M\,$ vanishes at $\,\rho=0$.
\,The different behavior of such diffusion for different values
of $\,\xi\,$ and $\,a\,$ has its origin in the singularity 
of $\,M\,$ at $\,\rho=0\,$ which requires different treatment in different
regimes. Up to the change of variables $\,x=\rho^{(2-\xi)/2}\,$ 
casting the generator $\,M\,$ into the form
\qq
M\ =\ {_1\over^4}(2-\xi)^2D'\,x^{1-\delta}\partial_x x^{\delta-1}\partial_x
\qquad{\rm for}\ \quad\delta\,=\,2(1-{_{1-a}\over^{2-\xi}})\label{BP} 
\qqq
and a time rescaling, the Markov process $\,\rho(t)\,$ may be identified 
with the well studied 
Bessel diffusion \cite{HBM}, a natural interpolation between processes 
describing the radial variable in the standard diffusion in  
dimensions $\,\delta$. Various analytic formulae may be then directly 
carried from that case to the present situation. 
\vskip 0.5cm

For $\,\xi=2\,$ corresponding to smooth velocities with velocity
differences linear in space, the particle dispersion PDF takes a
log-normal form \cite{CFKL,CKV}: \qq \CP(\rho_0,\rho;t)\ \ \
\propto\ \ \ee^{-{1\over4D_1't}\, (\ln(\rho/\rho_0)\,-\,\lambda
t)^2}\,\rho^{-1}\,, \label{KrI} \qqq where
$\,\lambda={d-4\wp\over1+2\wp}D_1'\,$ is the (biggest)
Lyapunov exponent. \,It is easy to see that in this case,
$\,\lim\limits_{\rho_0\to0}\,\CP(\rho_0,\rho;t)=\delta(\rho)$,
\,pointing to the deterministic nature of the flow.
Indeed, in velocities regular in space the trajectories  are
uniquely determined by their initial position and very close
fluid particles separate little in a fixed time interval. Nevertheless,
all moments of the pair dispersion behave exponentially in time 
and grow in the chaotic 
regime where $\,\lambda>0$, \,i.e. $\wp<{d/4}$, \,whereas sufficiently 
small (fractional) ones decrease when $\,\lambda<0$, i.e. $\wp>{d/4}$.
For the second moment, one obtains:
\qq \langle\,\rho^2(t)\,\rangle\ =\ {\rm
e}^{\,(2\lambda+4D_1')\,t}\, \rho_0^2\,. \label{sKR} \qqq Similar
behaviors persist at small times in the Kraichnan velocities with
$\,\xi>2\,$ and finite $\,L$, see \cite{LeJR1}. \vskip 0.5cm

The stochastic Lagrangian flow occurs in the non-Lipschitz version 
with $\,0<\xi<2\,$ of the Kraichnan model for weak compressibility 
$\,\wp<d/\xi^{2}$. \,In terms of the parameter $\,b={1-a\over2-\xi}
={2-\delta\over2}\,$ that will be frequently used below, the latter 
inequality means that $\,b<1$. \,In this region,
$\,\CP(\rho_0,\rho;t)=\ee^{-t\,M}(\rho_0,\rho)\,$ where $\,M\,$ is
taken with ``singular Neumann'' or reflecting boundary condition 
at $\,\rho=0$, see \cite{GV}, and \qq \lim\limits_{\rho_0\to0}\
\CP(0,\rho;t)\ \ \propto\ \ \rho^{a-\xi}\,t^{b-1}\,\,
\ee^{-{\rho^{2-\xi}\over(2-\xi)^2D_1'\,t}}\,. \label{KrII} \qqq 
Note the stretched-exponential form of the PDF.
In particular, one obtains for the second moment of the separation 
distance and large times:
\qq \langle\,\rho^2(t)\,\rangle\ =\ \CO(t^{2\over2-\xi})\,.
\label{KrR} \qqq This is the Kraichnan model
version of the 1926 Richardson law \cite{Rich} stating that in
developed turbulence the mean square dispersion grows
like $\,t^3$. \,Note that the Richardson behavior is reproduced in
the Kraichnan model for $\,\xi={4\over3}$. In the limit when the
initial trajectory separation $\,\rho_0\to0$, \, the power law
behavior (\ref{KrR}) extends to the entire time domain and
$\,\langle\rho^2(t)\rangle\propto\,t^{2\over2-\xi}$.  \vskip 0.5cm

In the non-Lipschitz strongly compressible version of the Kraichnan
model corresponding to $\,0<\xi<2\,$ and $\,\wp\geq d/\xi^{2}\,$ (or
$\,b\geq1$), \qq \CP(\rho_0,\rho;t)\ =\ \CP^{\rm
reg}(\rho_0,\rho;t)\ +\ p(\rho_0;t)\,\delta(\rho)\label{scwd}\qqq
with the regular part of the PDF $\,\CP^{\rm reg}(\rho_0,\rho;t)
=\ee^{-t\,M}(\rho_0,\rho)\,$ where $\,M\,$ is taken with
``singular Dirichlet'' or absorbing boundary condition at $\,\rho=0$, 
\,see \cite{GV}, and with \qq p(\rho_0;t)\ =\ 1-\gamma\left(b,
{_{\rho_0^{2-\xi}}\over^{(2-\xi)^2
D_1'\,t}}\right)\,\Gamma(b)^{-1}\,, \qqq where $\
\gamma(b,x)\equiv\int\limits_0^xy^{b-1}\,{\rm e}^{-y}\,dy\ $ is
the incomplete gamma-function. When $\,\rho_0\to0\,$ then the
regular part of the PDF tends to $\,0$, \,whereas $\,p(\rho_0;t)\,$
tends to $\,1$. \,We recover this way the deterministic behavior
(\ref{deter}), with trajectories in fixed velocity realizations
determined by their initial position. \,The presence of the term
proportional to $\,\delta(\rho)\,$ at finite $\,\rho_0\,$ signals
that the trajectories starting at different initial positions
coalesce with positive probability \cite{GV}. The time growth of
the mean distance square dispersion is different here: \qq
\langle\,\rho^2(t)\,\rangle\ =\ \CO(\rho_0^{1-a}\,t^{{2
\over2-\xi}-b}) \label{tg} \qqq with a logarithmic correction at
$\,\wp=d/\xi^{2}\,$ i.e.\,\,at $\,b=1$.

\subsection{\ Exit time}

\noindent In the Kraichnan model, the exit time PDF
$\,\CQ(t;\rho_0,\rho_1)\,$ may also be directly controlled using
the kernels $\,\ee^{-t\,M_{D}}(\rho_0,\rho)\,$ where $\,M_{D}\,$
denotes the generator $\,M\,$ of (\ref{M}) with the Dirichlet
boundary condition at $\,\rho_1\,$ (in addition to the appropriate
condition at the origin when $\,\rho_1>\rho_0$). This is due
to the Markov property of the stochastic process $\,\rho(t)$.
\,Since the exit times have not been discussed in the context
of the Kraichanan model, we shall use the occasion to provide more 
details, essentially translated form the Bessel diffusion case.  
\,The PDF of the time of exit through $\,\rho_1\,$ is given by 
\qq \CQ(t;\rho_0,\rho_1)\ =\ -\partial_{n(\rho_1)}
\,\ee^{-t\,M_{D}}(\rho_0,\rho)\,, \label{texII} \qqq where
$\,\partial_{n(\rho_1)}=\pm\,D_1'\,\rho^{a}\partial_{\rho}
\,\rho^{\xi-a}|_{\rho=\rho_1}\,$ plays here the same role as the normal
derivative in the classical potential theory. The sign is that of
$\,(\rho_1-\rho_0)$. \,This PDF does not
have to integrate to $\,1$. \,The eventual missing probability
corresponds to events where $\,\rho(t)\,$ stays forever in the
open interval $\,(0,\rho_1)\,$ or $\,(\rho_1,\infty)\,$ or when it
gets absorbed at the origin, see below. We shall assign the
infinite value of the exit time to such events. The averages of
the powers of the exit time $\,t\,$ over the realizations with
$\,t<\infty\,$ may be expressed by the kernels of the inverse
powers of operator $\,M_{D}$: \qq
\langle\,t^n\,1_{\{t<\infty\}}\,\rangle\ = \int\limits_0^\infty
t^n \,\CQ(t;\rho_0,\rho_1)\,dt \ =\
-\,n!\,\partial_{n(\rho_1)} M_{D}^{-n-1}(\rho_0,\rho;t)\,,
\label{unc} \qqq
where by $\,1_{\{A\}}\,$ we denote the characteristic
functions of the events satisfying the condition $\,A$.
\,In particular, the probability that the exit time
is finite \qq \langle\,1_{\{t<\infty\}}\,\rangle\ =
\int\limits_0^\infty \,\CQ(t;\rho_0,\rho_1)\,dt \ =\
-\partial_{n(\rho_1)}M_{D}^{-1}(\rho_0,\rho;t)\,. \label{tpf} \qqq
The expectations (\ref{unc}) may be obtained from the
characteristic function \qq \langle\,\ee^{i\omega
t}\,1_{\{t<\infty\}}\,\rangle\ =\ -\partial_{n(\rho_1)}
(M_{D}-i\omega)^{-1}(\rho_0,\rho_1;t) \label{shgf} \qqq involving
the resolvent kernel of $\,M_{D}$. We may also consider the
averages conditioned on the exit times being finite:\qq
{\langle\,t^n\,1_{\{t<\infty\}}\,\rangle\over
\langle\,1_{\{t<\infty\}}\,\rangle}\ \equiv\
\langle\,t^n\,\rangle_{_c}\,, \qquad{\langle\,\ee^{i\omega
t}\,1_{\{t<\infty\}}\,\rangle\over
\langle\,1_{\{t<\infty\}}\,\rangle} \ \equiv\
\langle\,\ee^{i\omega t}\,\rangle_{_c}\,. \label{shcf} \qqq \vskip
0.5cm

For $\,\xi=2\,$ and $\,L=\infty$, \,i.e.\,\,in the smooth version 
of the Kraichnan model,
the kernel $\,\ee^{-t\,M_{D}}(\rho_0,\rho)\,$ is easily calculable by
the image method:
\qq \ee^{-t\,M_{D}}(\rho_0,\rho)\ =\ \,{_1\over^{\sqrt{4\pi
D_1't}\,\rho}}
\Big(\ee^{-{1\over4D_1't}\,(\ln(\rho/\rho_0)\,-\,\lambda t)^2}
-\,\ee^{{\lambda\over D_1'}\ln(\rho_1/\rho_0)\,-\,{1\over4D_1't}\,
(\ln(\rho\rho_0/\rho_1^2)-\lambda t)^2}\Big)\,.\qqq For the exit
time PDF, one obtains: \qq \CQ(t;\rho_0,\rho_1)\,&=&\,{_{|\ln(\rho_1
/\rho_0)|}\over^{\sqrt{4\pi D_1'\,t^3}}}\,\ee^{-{1\over{4 D_1' t}}
(\ln(\rho_1/\rho_0)-\lambda t)^2}\cr &=&\,{_{|\ln(\rho_1/\rho_0)|}
\over^{\sqrt{4\pi D_1'\,t^3}}}\,\ee^{{\lambda\,\ln(\rho_1/\rho_0)
\over{2 D_1'}}}\,\ee^{-{1\over{4 D_1' t}}\ln^2(\rho_1/\rho_0)\,-\,
{\lambda^2\over{4 D_1'}}t}\,. \label{tpdf} \qqq
Note that the PDF depends only on $\,{\rho_1\over\rho_0}\,$
so that the corresponding Lagrangian flow is Lyapunov in our
terminology, see (\ref{lyap}). The total mass is given by the 
simple expression: 
\qq \int\limits_0^\infty \CQ
(t;\rho_0,\rho_1)\,dt\ =\ \cases{\ \hbox to 2.7cm{$1$\hfill}{\rm
if}\qquad \lambda\,\ln(\rho_1/\rho_0) \,\geq\,0\,,\cr \ \hbox to
2.7cm{$\left({{\rho_1}\over{\rho_0}}\right)^{{\lambda\over D_1'}}
$\hfill} {\rm if}\qquad\lambda\,\ln(\rho_1/\rho_0)\,\leq\,0\,.}
\qqq
The missing probability corresponds for $\,\rho_1>\rho_0\,$
and the negative Lyapunov exponent to the events
when $\,\rho(t)\,$ stays forever in the open interval
$\,(0,\rho_1)\,$ (the pairs of trajectories remain always close).
For $\,\rho_1<\rho_0\,$ and the positive Lyapunov exponent, it 
represents the events when $\,\rho(t)\,$ stays forever in the
interval $\,(\rho_1,\infty)\,$ (the pairs of trajectories never 
come close).
According to the characterization from the previous subsection,
see conditions (\ref{trans}), the Lagrangian flow is locally 
separating and locally
transient if $\,\lambda>0$, i.e. $\wp<{d/4}$\, and it is locally trapping
and locally recurrent
if $\,\lambda<0$, i.e. $\wp>{d/4}$. \,Finally, when $\,\lambda=0$, \,i.e. $\wp=d/4$,
\,it is locally separating and locally recurrent. \,Similar properties of the
exit time PDF hold for $\,\xi>2\,$ and finite $\,L\,$ asymptotically
at short distances.
\vskip 0.4cm

The exit time PDF (\ref{tpdf}) vanishes with all derivatives at
$\,t=0\,$ and it has an exponentially decaying tail at large
$\,t$. For the exit time conditional moments, one obtains:
\qq
\langle\,t^n\,\rangle_{_c}\,&=&\,{_{|\lambda|}\over^{\sqrt{\pi
D_1'}}}
\,\left({_{|\ln(\rho_1/\rho_0)|}\over^{|\lambda|}}\right)^{n+{_1\over^2}}
\ee^{{|\lambda\,\ln(\rho_1/\rho_0)|\over2D_1'}}\,K_{|n+{_1\over^2}|}
\left({_{|\lambda\,\ln(\rho_1/\rho_0)|}\over^{2D_1'}}\right)\cr
&=&\,\sum\limits_{k=0}^{|n-{_1\over^2}|-{_1\over^2}}{_{(|n-{_1\over^2}|
-{_1\over^2}+k)!}\over^{k!\,(|n-{_1\over^2}|
-{_1\over^2}-k)!}}\,\left({_{D_1'}\over^{\lambda^2}}\right)^k\,
\left({_{|\ln(\rho_1/\rho_0)|}\over^{|\lambda|}}\right)^{n-k},
\label{tns}\qqq where the first expression with the Bessel
function holds for all real $n$ and the second one for integer
$\,n$. \,The unconditioned moments diverge for $\,n>0\,$ if
$\,\lambda\,\ln(\rho_1/\rho_0)<0\,$ due to the finite probability
of infinite exit times. For the conditional characteristic
function one obtains: \qq \langle\,\ee^{i\omega t}\,\rangle_{_c}\
=\ \left({_{\rho_1}\over^{\rho_0}}\right)^{\pm{\lambda\over D_1'}
\Big({1\over2}\,-\,\sqrt{{1\over4}-i{D_1'\omega\over\lambda^2}}
\Big)}, \qqq where the square root is taken with the positive real
part and the sign is that of $\,\lambda\,\ln(\rho_1/\rho_0)$.
\,The decay at large $\,|\omega|\,$ and the analyticity properties
of the characteristic function reflect the behavior of the exit
time PDF at small and large $\,t$. \,In particular, the decay
$\,\sim\ee^{-|\ln(\rho_1/\rho_0)|\sqrt{|\omega|/D'_1}}\,$ of the
characteristic function along the positive imaginary axis of
$\,\omega\,$ signals the behavior
$\,\sim\ee^{-{1\over4D'_1\,t}\ln^2(\rho_1/\rho_0)}\,$ of
$\,\CQ(t;\rho_0,\rho_1)\,$ for small $\,t\,$ and the
singularity at $\,\omega=-i{\lambda^2\over4D'_1}\,$ indicates the
presence of the tail $\,\sim\ee^{-{\lambda^2\over4D'_1}t}\,$ for
large $\,t$, \,in agreement with Eq.\,\,(\ref{tpdf}). Note
non-Gaussian character of the large fluctuations of $\,t\,$
and of $\,t^{-1}$.
\vskip 0.4cm

In the non-Lipschitz version of the Kraichnan model with $\,0<\xi<2$, 
\,the resolvent kernel $\,(M_{D}-i\omega)^{-1}
(\rho_0,\rho)\,$ may still be easily calculated in a closed form.
Let us start from the case with $\,\rho_1>\rho_0$.
On the interval $\,[0,\rho_1]$,
\qq &&(M_{D}-i\omega)^{-1}(\rho_0,\rho)\ =\
{_1\over^{D_1'}}\,\CW^{-1}\,\rho^{a-\xi} \,f_\mp(\rho_1)^{-1}\cr
&&\hspace{2cm}\cdot\ \cases{\hbox to
7.3cm{$\,f_\mp(\rho_0)\,(f_\pm(\rho)
\,f_\mp(\rho_1)\,-\,f_\mp(\rho)\,f_\pm(\rho_1))$\hfill}{\rm
for}\quad\rho_0\leq \rho\,,\cr \hbox to
7.3cm{$\,f_\mp(\rho)\,(f_\pm(\rho_0)
\,f_\mp(\rho_1)\,-\,f_\mp(\rho_0)\,f_\pm(\rho_1))$\hfill}{\rm
for}\quad\rho_0\geq \rho\,,} \label{MDo} \qqq where the upper sign
pertains to the weakly compressible $\,\wp<{d\over\xi^2}\,$ region
and the lower one to the strongly compressible one
$\,\wp\geq{d\over\xi^2}$, \,with $\,f_\pm\,$ giving
two independent solutions of the eigenfunction equation
$\,(M-i\omega)f=0\,$ expressed by the Bessel functions: \qq
f_\pm(\rho)\ =\ \rho^{1-a\over2}\,J_{\pm b}\left({_2
\over^{2-\xi}}\sqrt{{_{i\omega}\over^{D_1'}}\,\rho^{2-\xi}}\right)
\qqq  and with \qq \CW\ =\ \rho^a\left(f_\pm(\rho)\partial_\rho
f_\mp(\rho)\,-\,f_\mp(\rho)
\partial_\rho f_\pm(\rho)\right)
\label{Wron} \qqq standing for their $\rho$-independent Wronskian.
The eigenfunction $\,f_+\,$ ($\,f_-$) \,satisfies the singular
Dirichlet (Neumann) condition at the origin imposed by the limit
when the trajectory noise is turned off for weak (strong)
compressibility, see \cite{GV}. It follows that \qq
\langle\,\ee^{i\omega t}\,1_{\{t<\infty\}}\,\rangle\ =\
-\partial_{n(\rho_1)}(M-i\omega)^{-1}(\rho_0,\rho_1)\ =\
{{f_\mp(\rho_0)}\over{f_\mp(\rho_1)}}\,. \label{gfo} \qqq For
$\,\omega\to0\,$ the eigenfunction $\,f_-$ reduces to a constant
whereas $\,f_+$ becomes proportional to $\,\rho^{1-a}$ so that \qq
\langle\,1_{\{t<\infty\}}\,\rangle\ =\ \cases{\,\hbox to
2cm{$1$\hfill}{\rm for}\quad\wp<{d\over\xi^2}\,,\cr\,\hbox to
2cm{$\left({\rho_0\over\rho_1}\right)^{^{1-a}}$\hfill}{\rm
for}\quad\wp\geq{d\over\xi^2}\,. }\qqq Hence the exit time is
almost surely finite in the weakly compressible regime whereas it
is infinite with positive probability that depends only on $\,{\rho_1
\over\rho_0}\,$ in the strongly compressible
regime where the process $\,\rho(t)\,$ is absorbed at the origin
with the complementary probability. Such absorption corresponds
to the coalescence of pairs of trajectories. We conclude that
the Lagrangian flow is locally separating for $\,\wp\leq d/\xi^2\,$ 
and locally trapping for $\,\wp>d/\xi^2$.
\vskip 0.4cm

For the conditional characteristic function,
one obtains \qq \langle\,\ee^{i\omega t}\,\rangle_{_c}\ =\
\left({\rho_0\over\rho_1}\right)
^{\mp{a-1\over2}}\,{{J_{\mp b}\left({2\over{2-\xi}}
\sqrt{{_{i\omega}\over^{D_1'}}\,\rho_0^{2-\xi}}\right)}\over
J_{\mp b}\left({2\over{2-\xi}}\sqrt{{_{i\omega}
\over^{D_1'}}\,\rho_1^{2-\xi}}\right)}\,. \label{rhs} \qqq The
moments of the exit times may be  derived from this expression by
expanding the right hand side in powers of $\,\omega$. In
particular, one obtains for the conditional average of the exit
time the result: \qq \langle\,t\,\rangle_{_c}\ =\
{_{(\rho_1/\rho_0)^{^{2-\xi}}-1}
\over^{(2-\xi)(2-\xi\mp(1-a))D_1'}}\,\rho_0^{2-\xi} \label{dbt}
\qqq which reproduces in the $\,\xi\to2\,$ limit the $\,n=1\,$
version of Eq.\,\,(\ref{tns}). The decay $\,\propto
\ee^{-\CO(\sqrt{|\omega|})}\,$ of the absolute value of the right
hand side of Eq.\,\,(\ref{rhs}) at large positive or negative
$\,\omega\,$ guarantees that the exit time PDF
$\,\CQ(t;\rho_0,\rho)\,$ is smooth. Since it is zero for negative
$\,t$, it must vanish with all derivatives at $\,t=0$. \,More
exactly, the decay $\,\sim\ee^{-b_1\sqrt{|\omega|}}\,$ of the
characteristic function (\ref{rhs}) along the positive imaginary
axis of $\,\omega$, \,with $\,b_1=2(2-\xi)^{-1}(D_1')^{-1/2}
(\rho_1^{(2-\xi)/2}-\rho_0^{(2-\xi)/2})$, \,signals the behavior
$\,\sim\ee^{-{b_1^2\over4t}}\,$ of $\,\CQ(t;\rho_0,\rho_1)\,$ for
small $\,t$. The analyticity properties of the right hand side of
(\ref{rhs}) imply the exponential decay $\,\sim\ee^{-b_2t}\,$ of
$\,\CQ(t;\rho_0,\rho_1)\,$ for large $\,t$, \,with the rate
$\,b_2={_1\over^4}(2-\xi)^2D_1'\rho_1^{\xi-2}x_0^2$ \,where
$\,x_0\,$ is the (real) zero of $\,J_{\mp b}(z)\,$
closest to the origin. In this respect, the exit time PDF
$\,\CQ(t;\rho_0,\rho_1)\,$ behaves similarly for the weak
and for the strong compressibility, the main difference between
the two cases consisting in the missing probability in the latter case.
\vskip 0.4cm

For $\,\rho_1<\rho_0$, the statistics of the time of exit through
$\,\rho_1$ is related to the resolvent kernel of the generator
$\,M_{D}\,$ on the interval $\,[\rho_1,\infty)$. \,For $\,0<\xi<\,2$,
\,the latter is given by a formula like (\ref{MDo}) but with
the overall minus sign, the cases $\,\rho_0\leq\rho\,$ and 
$\,\rho_0\geq\rho\,$ interchanged, and the functions 
$\,f_{\mp},\,f_{\pm}\,$ replaced by the Hankel functions 
\qq
f^{(i)}(\rho)\ =\
\rho^{1-a\over2}\,H^{(i)}_{b}\left({_2
\over^{2-\xi}}\sqrt{{_{i\omega}\over^{D_1'}}\,\rho^{2-\xi}}\right)
\label{hae} \qqq 
for $\,i=1,2$, \,respectively.
The square root in the argument of the Hankel functions
should be taken with the positive imaginary part so that it is the
eigenfunction $\,f^{(1)}\,$ which has a stretched exponential
decay for large $\,\rho$. For the characteristic function of the
exit time, Eq.\,\,(\ref{shgf}) gives: \qq
\langle\,\ee^{i\omega t}\,1_{\{t<\infty\}}\,\rangle\ =\
{f^{(1)}(\rho_0)\over f^{(1)}(\rho_1)}\,.\label{gfp} \qqq When
$\,\omega\to0$, \,the eigenfunction $\,f^{(1)}\,$ becomes
proportional to $\,\rho^{1-a}\,$ if $\,a>1$, i.e.\,\, if
$\,b< 0\,$ or $\,\wp<{d-2\over2\xi}+{1\over2}\,$ and to a constant
if $\,a\leq1$, i.e.\,\,if $\,b\geq0\,$ or $\,\wp\geq{d-2\over2\xi}
+{1\over2}$. \,It follows that \qq
\langle\,1_{\{t<\infty\}}\,\rangle\ =\ \cases{\,\hbox to 2
cm{$\left(\rho_1\over\rho_0\right)^{a-1}$\hfill}{\rm
for}\quad\wp<{d-2\over2\xi}+{1\over2}\,,\cr \,\hbox to 2
cm{$1$\hfill}{\rm for}\quad\wp\geq{d-2\over2\xi}+{1\over2}\,.} \qqq For
$\,\wp<{d-2\over2\xi}+{1\over2}\,$ (i.e.\,\, for $\,a>1$), the process
$\,\rho(t)\,$ remains forever in the interval
$\,(\rho_1,\infty)\,$ with probability
$\,1\,-\,\left(\rho_1\over\rho_0\right)^{a-1}\,>\,0\,$ whereas for
$\,\wp\geq{d-2\over2\xi}+{1\over2}\,$ it exits through $\,\rho_1\,$
almost surely corresponding, respectively, to a locally transient 
and a locally recurrent Lagrangian flow.
\vskip 0.4cm

The conditional characteristic function is given by the expression \qq
\langle\,\ee^{i\omega t}\,\rangle_{_c}\ =\
\Big({\rho_0\over\rho_1}\Big)^{^{|1-a|\over2}}\,{H^{(1)}_{b}
\left({2
\over{2-\xi}}\sqrt{{_{i\omega}\over^{D_1'}}\,\rho_0^{2-\xi}}\right)\over
H^{(1)}_{b}\left({2
\over{2-\xi}}\sqrt{{_{i\omega}\over^{D_1'}}\,\rho_1^{2-\xi}}\right)}\,.
\label{Hcha} \qqq Again, its absolute value decays as
$\,\ee^{-\CO(\sqrt{|\omega|})}\,$ for large $\,|\omega|\,$
implying that $\,\CQ(t;\rho_0,\rho_1)\,$ is smooth and
vanishes with all derivatives at the origin. More exactly, the
decay $\,\sim\ee^{-b_1\sqrt{|\omega|}}\,$ of (\ref{Hcha}) along
the positive imaginary axis, where $\,b_1$ is as for
$\,\rho_1>\rho_0\,$ but with $\,\rho_0$ and $\,\rho_1\,$
interchanged, implies again the behavior
$\,\sim\ee^{-{b_1^2\over4t}}\,$ of the exit time PDF
$\,\CQ(t;\rho_0,\rho_1)\,$ for small $\,t$. \,Since
$\,H^{(1)}_b(z)\,$ is a combination of $\,z^{\pm b}\,$ with
coefficients that are entire functions of $\,z^2$ (for non-integer
$\,b$), it follows that $\,\langle\,\ee^{i\omega t}\,\rangle_{_c}$
has the $n^{\rm th}$ derivative over $\,\omega\,$ at the origin if
(and only if) $\,n<|b|$. \,That implies that for
$\,\rho_1<\rho_0$ the exit time PDF
$\,\CQ(t;\rho_0,\rho_1)\,$ has a power decay for large $\,t$,
\,unlike for $\,\rho_1>\rho_0$ where it decayed exponentially:
the large deviations of $\,t\,$ are even more non-Gaussian
in this case.

\vskip 0.5cm

For both $\,\rho_1>\rho_0\,$ and $\,\rho_1<\rho_0$, \,the Kraichnan
model exit time
PDF has for $\,0<\xi<2\,$ the scaling property: \qq \mu\,\CQ(\mu
t;\mu^{1\over2-\xi}\rho_0,\mu^{1\over2-\xi}\rho_1)\ =\
\CQ(t;\rho_0,\rho_1)\,. \qqq It follows that the moments of the exit time
$\,\langle\, t^n\,\rangle\,$ are
proportional to $\,\rho_0^{n(2-\xi)}\,$ if $\,{\rho_1\over\rho_0}\,$ is
kept constant. The same scaling implies the  behavior (\ref{Rich})
that characterizes what we have called the Richardson flows.
This behavior may be also directly inferred from Eqs.\,\,(\ref{gfo}) 
and (\ref{gfp}).
\vskip 0.5cm

\begin{figure}[t]
\begin{center}
\mbox{\hspace{0.0cm}\epsfig{file=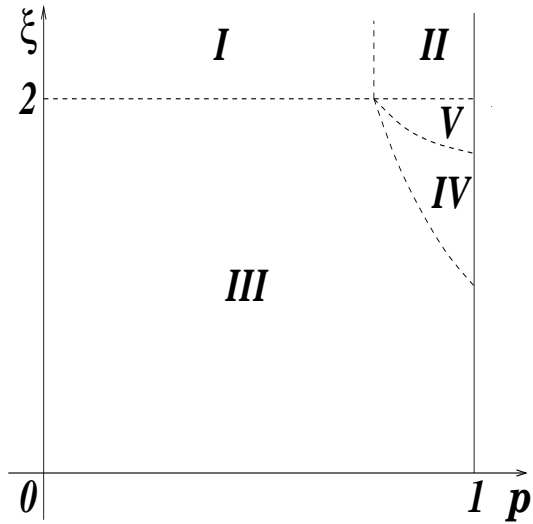,height=7cm,width=7cm}}
\end{center}
\vspace{0.3cm}
\caption{Phase diagram of the Lagrangian flow for 
the three-dimensional Kraichnan model.}
\label{fig1}
\end{figure}
\vskip 0.5cm

One may summarize the properties of the Lagrangian flow in the Kraichnan
model in the phase diagram, drawn in Fig.~\ref{fig1} for three dimensions,
with five phases that we list with their characteristics (assuming
for $\,\xi>2\,$ finite $\,L$):
\vskip 0.5cm

\ \hbox to 0.9cm{{\bf I}.\hfill}deterministic, Lyapunov, locally separating, 
locally transient,

\ \hbox to 0.9cm{\hfill}for $\,\xi>2\,$ and $\,\wp<d/4$\,;
\vskip 0.5cm

\ \hbox to 0.9cm{{\bf II}.\hfill}deterministic, Lyapunov, locally 
trapping, locally recurrent,

\ \hbox to 0.9cm{\hfill}for $\,\xi>2\,$ and $\,\wp>d/4$\,;
\vskip 0.5cm

\ \hbox to 0.9cm{{\bf III}.\hfill}stochastic, Richardson, locally separating, 
locally transient,

\ \hbox to 0.9cm{\hfill}for $\,0<\xi<2\,$ and $\,\wp<{d-2\over2\xi}
+{1\over2}$\,;
\vskip 0.5cm

\ \hbox to 0.9cm{{\bf IV}.\hfill}stochastic, Richardson, locally 
separating, locally recurrent,

\ \hbox to 0.9cm{\hfill}for $\,0<\xi<2\,$ and $\,{d-2\over2\xi}+{1\over2}<\wp
<d/\xi^2$\,;
\vskip 0.5cm

\ \hbox to 0.9cm{{\bf V}.\hfill}deterministic, Richardson, locally trapping, 
locally recurrent,

\ \hbox to 0.9cm{\hfill}for $\,0<\xi<2\,$ and $\,d/\xi^2<\wp$\,.
\vskip 0.4cm

\noindent These phases were essentially enumerated in \cite{GV}
(with little stress put on the difference between phase {\bf III}
and {\bf IV}). The characterization described above is closely related
to the description of the phase diagram in \cite{LeJR1}, see also \cite{EVDE1}.
The notable difference is that, in order to characterize the phases 
in the non-Lipschitz case, refs.\,\cite{LeJR1,EVDE1} used the dichotomic behaviors 
of the time of exit through $\,\rho_1\,$ in two limits: when $\,\rho_1\to0\,$ 
with $\,\rho_0\,$ kept constant and when $\,\rho_0\to0\,$ with $\,\rho_1\,$ 
kept constant. Such behaviors enter the standard classification \cite{Feller,Breiman}
of the one-dimensional diffusion $\,\rho(t)\,$ on the half-line $[0,\infty[$ 
with $\,\rho=0\,$ being, for $\,\xi>2$, \,a natural boundary and,
for $\,\xi<2$, \,an entrance boundary in the weakly compressible phase III,
a regular boundary in the intermediate phase IV and an exit boundary in 
the strongly compressible phase V. The use in the present paper
of the small $\,\rho_0\,$ behavior of the exit time PDF at fixed 
$\,{\rho_1\over\rho_0}\,$ in order to characterize the phases was motivated
by the fact that such behaviors were both more amenable to analytic arguments 
in the presence of temporal correlations of velocities and more accessible 
to numerical simulations.
\vskip 0.4cm

As noticed in \cite{EVDE1}, see also \cite{LeJR2}, 
the solution for the PDF $\,\CP(\rho_0,\rho;t)\,$ corresponding
to the singular Dirichlet boundary condition for $\,M\,$
and coalescent trajectories, which pertains only to phase {\bf V}
if the flow is defined by adding and removing small noise, sets
in already in region {\bf IV} if we add no noise but first smoothen out
the velocity fields at short distances and subsequently remove
the smoothing. Physically, the first procedure corresponds to
the vanishing Prandtl and the second one to the infinite Prandtl
numbers. For $\,{d-2\over2\xi}+{1\over2}<\wp<{d\over\xi^2}\,$ and well
tuned Prandtl numbers, one may also obtain intermediate solutions
that correspond to a ``sticky'' behavior of fluid particles \cite{H}.
The different limiting procedures give then rise to different boundary
conditions that may be imposed on the generator $\,M\,$ of \,Eq.\,\,(\ref{M})
in the situation when $\,\rho=0\,$ is a regular boundary 
\cite{Feller,Breiman,EVDE2}.   
\vskip 0.5cm

\nsection{Gaussian velocity ensembles with temporal correlations}
\label{sec:GVTC}

\noindent The temporal decorrelation of the Kraichnan velocities is a
simplifying feature that is quite unphysical since realistic
turbulent velocities are correlated at different times. In the
present paper we attempt to study the effect of temporal
correlation of velocities on the behavior of the dispersion of a
pair of particles in simplest ensembles of velocities with such
correlations built in. More specifically, we shall consider the
Gaussian ensembles of $d$-dimensional velocities with mean zero
and covariance \qq \Big\langle v^i(t,{\bm r})\, v^j(t',{\bm
r}')\Big\rangle \ =\ D_2\int {\rm
e}^{-|t-t'|D_3\,k_{_L}^{2\beta}}\,\,{{{\rm e}^{\,i\, {\bm
k}\cdot({\bm r}-{\bm r}')}}
\over{k_{_L}^{d+2\alpha}}}\,\,P^{ij}({\bm k},\wp) \,\,{{d{\bm
k}}\over{(2\pi)^d}}\,. \label{vc} \qqq There are three parameters
in (\ref{vc}) not related to the choice of units: the spatial
H\"older exponent $\,\alpha$, \,that we shall restrict to the
interval $\,(0,1)$, \,the temporal exponent $\,\beta\,$ taken positive, 
\,and the compressibility degree
$\,\wp\in[0,1]\,$. \,Besides, there are three dimensionful
parameters: $\,D_2$ of dimension $\,length^{2(1-\alpha)}/time^2$,
$\,D_3$ of dimension $\,length^{2\beta}/time$, \,and the integral
length scale $\,L$. \,Similarly as in the Kraichnan ensemble, $L$
may be taken to infinity for correlation functions of differences
of velocities $\,\Delta{\bm v}(t,\brho)\,$ whose statistics
becomes scale invariant in this limit. The correlation time
$\,\tau_c(\rho)\,$ of the velocity differences in ensembles given by
Eq.\,\,(\ref{vc}) is equal to $\,D_3^{-1}\rho^{2\beta}\,$
whereas the variance $\,\langle(\Delta\bv(t,\brho))^2\rangle\equiv
\Sigma(\rho)^2\,$ behaves as $\,D_2\rho^{2\alpha}$. \vskip 0.5cm

We shall be looking at the statistics of the 2-particle separation
either using equation (\ref{rho}) with the ensemble (\ref{vc})
governing the {\bf Eulerian} velocities, \,or using
Eq.\,\,(\ref{rho1}) with the ensemble (\ref{vc}) describing the
{\bf quasi-Lagrangian} velocities. \,It should be stressed that
the two choices lead to two different models of Lagrangian flow.
They exhibit different behaviors even for incompressible
velocities where the equal-time velocity statistics of the
Eulerian and quasi-Lagrangian velocities coincide. This should be
contrasted with the situation in the Kraichnan model where the
Eulerian and the quasi-Lagrangian velocities had the same all-time
statistics so that it did not matter which one was modeled with
the Gaussian ensemble (\ref{vcKr}). That the situation is
different in the presence of temporal correlations is due to the
manner in which sweeping by large scale eddies is taken into
account. The 2-particle separation $\,\brho\,$ involves only differences
$\,\Delta\bv^{qL}(t,\brho)\,$ of the quasi-Lagrangian velocities,
see (\ref{rho1}). The statistics of such differences has a regular
$\,L\to\infty\,$ limit if we use the Gaussian ensemble (\ref{vc})
for the quasi-Lagrangian velocities. In this case the sweeping by
scale $L$ eddies does not effect the pair dispersion. On the other
hand, $\,\brho\,$ cannot be expressed in terms of the Eulerian
velocity differences only, due to the dependence on the reference
trajectory $\,\bR(t)$, see (\ref{rho}). As a result, if we
substitute the ensemble (\ref{vc}) for the Eulerian velocities,
the dispersion statistics is still effected by the scale $L$
sweeping and behaves in a singular way in the limit
$\,L\to\infty$. This singularity modifies also the short time
behavior of the pair dispersion at fixed $L$ in certain regimes,
as we shall discuss below. The use of the synthetic ensemble
(\ref{vc}) to describe turbulent velocities is in any case an
approximation. It seems to render better the Lagrangian features
of real turbulence if used to model the quasi-Lagrangian
velocities. In this case the large scale sweeping influences only
the single particle statistics, but not the pair dispersion. We
shall limit ourselves to this situation throughout most of this
paper, dropping the subscript ``${qL}$'' on the velocities. The
exception is Sect.\,\ref{sec:EUL} where we discuss what happens
when the Gaussian ensemble (\ref{vc}) is used to model the
Eulerian velocities. 

\begin{figure}
\begin{center}
\vspace{0.3cm}
\mbox{\hspace{0.0cm}\epsfig{file=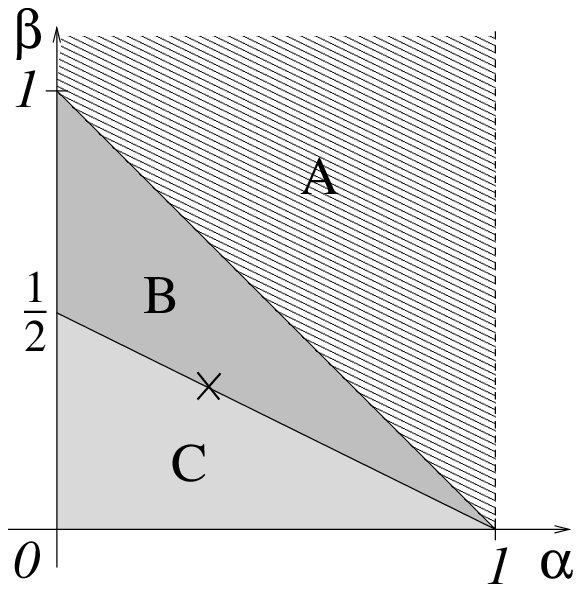,height=6.5cm,width=6.5cm}}
\end{center}
\vspace{0.3cm}
\caption{Phase diagram of the three different regimes 
of Lagrangian flow in the time-correlated velocities discussed here.
The exponent $\alpha$ is the spatial H\"older exponent and the exponent 
$\beta$ controls the behavior of correlation time {\it vs} scale.}
\label{fig2}
\end{figure}
\vskip 0.6cm

The first idea about the Lagrangian flow in the ensembles
(\ref{vc}) of quasi-Lagrangian velocities may be gained by
comparing the correlation time $\,\tau_c(\rho)\,$ to the eddy 
turnover time $\,\tau_e(\rho)=D_2^{-1/2}\rho^{1-\alpha}\propto
{\rho}/\Sigma(\rho)$. \,On the line
$\,\alpha+2\beta=1\,$ which, in particular, contains the
Kolmogorov point $\,\alpha=\beta ={1\over3}$ (see Fig.~\ref{fig2}), both
times have the same scale dependence. For $\,\alpha+2\beta<1\,$
(domain \,C\, on Fig.~\ref{fig2}), $\,\tau_c\,$ becomes much shorter than
$\,\tau_{e}$ at large scales and much longer at small ones.
We shall see that in this domain of parameters the pair dispersion
should be described at long scales (for $L=\infty)$ by the
Kraichnan model with rapidly decorrelating velocities and at short
scales by a model with velocities independent of time (frozen).
For $\,\alpha+2\beta>1$ (domains \,A\, and \,B\, on Fig.~\ref{fig2}), the
relation between the correlation times is reversed and we could
expect that the frozen model controls the large scale dispersion
and the decorrelated one governs the short scale behavior. A
similar picture underlined the phase diagram in a simple family of
scale-invariant shear flows \cite{MA}. We shall study
the asymptotics of the Lagrangian flow using scale
transformations. Such transformations induce a flow in the plane
of dimensional parameters of the model whose asymptotics is
controlled by fixed points, as in the field theory renormalization
group \cite{Amit}. The perturbative renormalization group has been
previously used in \cite{Ant} to analyze the related scalar
advection problem in the family (\ref{vc}) of velocity ensembles
around $\,(\alpha,\beta)=(-1,1)\,$ and there is some overlap of
those results with our conclusions. The much heavier analysis of
\cite{Ant} was concentrated, however, on the aspects of advection
related to finer details of the Lagrangian flow, see also
\cite{CFL}. Our point is that the analysis of the Lagrangian
dispersion may be performed, at least to at certain extent, in a 
straightforward and nonperturbative way. \vskip 0.5cm
\vskip 0.4cm

The Kraichnan and the frozen model may be viewed as special
limiting cases of the Gaussian velocity ensembles (\ref{vc}). The
first one, with $\,\xi=2(\alpha+\beta)$, is obtained when
$\,D_2,D_3\to\infty\,$ with $\,{2D_2\over D_3}\equiv D_1\,$ kept
constant as a consequence of the convergence \qq
D_3k_{_L}^{2\beta}\,\ee^{-|t|\,D_3k_{_L}^{2\beta}}\
\mathop{\longrightarrow}\limits_{D_3\to\infty}\ 2\,\delta(t)\,.
\qqq At $\,L=\infty$, \,existence of the limit for the correlation
functions of $\,\Delta v\,$ requires that $\,\xi<2$,
\,\,i.e.\,\,that $\,\alpha+\beta<1$. \,Note that the convergence
is fast at large wave number $\,k$, \,i.e.\,\,at small distances,
and for long times, but it becomes slow for short times and, if
$\,L=\infty$, \,at large distances. The frozen model is obtained
by taking $\,D_3\to0\,$ with $\,D_2={\rm const}$. In this case,
\qq \ee^{-D_3|t|\,k_{_L}^{2\beta}}\ \mathop{\longrightarrow}
\limits_{D_3\to0}\ 1\qqq but the convergence becomes slow at large
$\,k$, \,i.e.\,\,at small distances, and for long times. \vskip
0.5cm

Using the evolution equation (\ref{rho1}) for the trajectory
separation vector, we obtain for the mean rate of growth of the
square of pair dispersion: \qq {d\over
dt}\,\langle\,\rho^2(t)\,\rangle\ =\ 2\int\limits_0^t
\Big\langle\,\Delta\bv(t,{\bm\rho}(t))\cdot\Delta\bv(s,{\bm\rho}(s))
\,\Big\rangle\,\,ds\,. \label{nes} \qqq Let us start by a naive
mean-field-type approximate evaluation of the right hand side in 
the limit when $\,\rho_0\to0$. Such an evaluation should render correctly the
behavior of $\,\langle\,\rho^2(t)\,\rangle\,$ in the stochastic
regime. It is obtained by rewriting Eq.\,\,(\ref{nes}) as \qq {d\over
dt}\,\langle\,\rho^2\,\rangle\ =\ 2\,T\,
\,\,\Big\langle\,(\Delta\bv)^2(t,{\bm\rho})\,\Big\rangle\,,
\label{rew} \qqq where \qq T\ =\
{\int\limits_0^t\Big\langle\,\Delta\bv(t,{\bm\rho}(t))\cdot
\Delta\bv(s,{\bm\rho}(s))\,\Big\rangle\,\,ds \over
\Big\langle\,(\Delta\bv)^2(t,{\bm\rho}(t))\,\Big\rangle}\,,\qqq
has, if smaller than $\,t$, \,an interpretation of the correlation 
time of the Lagrangian
velocity difference $\,\Delta\bv(t,{\bm\rho}(t))$. \,We may try to
close Eq.\,\,(\ref{rew}) by assuming that $\,T\,$ depends on the
mean separation distance $\,\langle\rho^2\rangle^{1/2}\,$ the same
way as the correlations time $\,\tau_c(\rho)\,$ on $\,\rho\,$ if
$\,\tau_c(\rho)\,$ is smaller than $\,t\,$ or as $\,t\,$ otherwise:
\qq T\ \approx\ {\rm
min}\{\,\CO(\langle\rho^2\rangle^\beta)\,,\,t\,\}\,. \label{et}
\qqq Different domains in the space of parameters correspond to
different choices of the minimal value (\ref{et}) for $\,T$.
\,As for the other term on the right hand side of (\ref{rew}), we
shall again ignore the velocity dependence of $\,\rho(t)\,$
putting \qq \Big\langle\,(\Delta\bv)^2\,\Big\rangle\ \approx\
\CO(\langle\rho^2 \rangle^\alpha) \label{ln} \qqq Using the above
approximations for {\bf long times} and $L=\infty$, one obtains
from (\ref{rew}): \qq &&\hspace{0.6cm}\langle\,\rho^2(t)\,\rangle\
\ \approx \ \ \hbox to 3cm{$\CO(t^{2\over1 -\alpha})$\hfill}{\rm
for}\ \ \alpha+2\beta\geq1\ \ \ \hspace{1.6cm}
({\rm domains\ A\ and\ B})\,,\label{ltm1}\\
&&\hspace{0.6cm}\langle\,\rho^2(t)\,\rangle\ \
\approx\ \ \hbox to 3cm{$\CO(t^{1\over1
-\alpha-\beta})$\hfill}{\rm for}\ \ \alpha+2\beta\leq1\ \
\ \hspace{1.6cm}({\rm domain\ C})\,.
\label{ltm2}
\qqq
In the same way, we may estimate the {\bf short-time} behavior
in the limit $\,\rho_0\to0\,$ obtaining
\qq
&&\hspace{-0.55cm}\langle\,\rho^2(t)\,\rangle\ \ \approx
\ \ \hbox to 3cm{$\CO(t^{1\over1
-\alpha-\beta})$\hfill}{\rm for}\,\ \ \hbox to 4cm{$\alpha+\beta<1
\leq\alpha+2\beta$\hfill}({\rm domain\ B})\,,
\label{stm2}\\
&&\hspace{-0.55cm}\langle\,\rho^2(t)\,\rangle\ \ \approx \ \ \hbox
to 3cm{$\CO(t^{2\over1 -\alpha})$\hfill}{\rm for}\,\ \ \hbox to
4cm{$\alpha+2\beta\leq1$\hfill} ({\rm domain\ C})\,. \label{stm1}
\qqq Note that, in agreement with (\ref{et}), it is the smaller
exponent that is chosen for large times and the bigger one for
short times, a manifestation of a tendency of close trajectories
to stay close. The region $\,\alpha+\beta\geq1\,$ (domain A in
Fig.~\ref{fig2}) which escapes the short-time estimates has been
rigorously analyzed with the Gaussian ensemble (\ref{vc}) used 
to model the Eulerian velocities \cite{FK} and was conjectured to 
correspond to deterministic trajectories. We expect that also in the 
quasi-Lagrangian model the pair dispersion will concentrate in 
domain A at $\,\rho=0\,$ when $\,\rho_0\to0$. This is consistent 
with the divergence of the predicted power in the short-time 
Richardson law (\ref{stm1}) when $\,\alpha+\beta\,$ approaches 
$\,1\,$ from below (i.e.\,\,from the domain B in Fig.~\ref{fig2}). A similar 
divergence occurs in the weakly compressible Kraichnan model when 
we approach the Lipschitz regime $\,\xi>2\,$ from the non-Lipschitz 
one $\,\xi<2$. \,It signals there the passage from the power law 
to the exponential separation (\ref{sKR}) of trajectories. 
\vskip 0.5cm

\nsection{Scaling arguments} \label{sec:SA}

\noindent The main aim of this note is to substantiate further the above
conclusions based on the naive estimates (\ref{et}) and
(\ref{ln}). We shall also acquire an insight into the behavior of
general moments of the pair dispersion and of the exit time and
into the extent of the different Lagrangian flow regimes in the
ensembles given by Eq.\,\,(\ref{vc}). In the study of the long-
and short-time behavior of trajectories, it is convenient to
consider their rescaled versions $\,\bR_\mu(t)=\mu^{-\sigma}
\bR(\mu t)\,$ for appropriately chosen $\sigma$. \,Since \qq
{d\bR_\mu\over dt}\ =\ {\bm v}_\mu(t,\bR_\mu)\,, \qqq for the
rescaled velocity $\,\bv_\mu(t,\br)=\mu^{1-\sigma} \bv(\mu
t,\mu^\sigma\br)$, the path $\,\bR_\mu(t)$ is a Lagrangian
trajectory for $\,\bv_\mu$. \,There are special cases when the
rescaled velocity differences have in the $\,L\to\infty\,$ limit
the same distribution as the original ones for an appropriate (and
unique) choice of $\,\sigma$. This happens for $\,\sigma={1\over1-\alpha}\,$ 
both on the line $\,\alpha+2\beta=1\,$ and in the frozen model and
for $\,\sigma={1\over2-\xi}$ in the Kraichnan ensemble. \,We infer that 
in those cases the pair dispersion PDF $\,\CP\,$ and the exit time 
one $\,\CQ\,$ are scale-invariant: \qq
\mu^\sigma\,\CP(\mu^\sigma\rho_0,\mu^\sigma\rho;\mu t)\ =\
\CP(\rho_0,\rho;t)\,,\qquad \mu\,\CQ(\mu
t;\mu^\sigma\rho_0,\mu^\sigma\rho_1)\ =\ \CQ(t;\rho_0,\rho_1)\,.
\label{scp} \qqq In particular, in the stochastic regime, the pair
dispersion moments $\,\langle\m\rho(t)^n\m\rangle$, \,if finite in the
$\,\rho_0\to0\,$ limit, behave as $\,\CO(t^{n\sigma})\,$ for long
times and become proportional to $\,t^{n\sigma}\,$ for all times
when $\rho_0\to0$. These conclusions fail in the deterministic
regime, as we have already noticed in the Kraichnan model, see
Eq.\,\,(\ref{tg}). In all regimes, the exit time moments
$\,\langle\,t^n\,1_{\{t<\infty\}}\,\rangle\,$ (if finite) are 
proportional to $\,\rho_0^{n/\sigma}\,$ if $\,{\rho_1\over\rho_0}
\equiv\gamma\,$ is kept constant. \vskip 0.5cm

Out of the line $\,\alpha+2\beta=1$, \,the scale invariance of the
Lagrangian dispersion is broken but in a predictable way, as we
shall see. The crucial observation is that the rescaled velocities
$\,\bv_\mu(t,\br)\,$ are distributed with the 2-point function
(\ref{vc}) with $\,D_i$ replaced by $\,D_{i,\mu}$ and $\,L\,$ by
$\,L_\mu$, \,where \qq
D_{2,\mu}\,=\,\mu^{2[1-(1-\alpha)\sigma]}D_2\,\qquad D_{3,\mu}\,
=\,\mu^{1-2\beta\sigma}\,D_3\,,\qquad L_\mu=\mu^{-\sigma}L\,.
\label{resc} \qqq For $\,\mu\,$ tending to infinity or to zero
(i.e.\,\,when exploring the long-time or the short-time behavior
of the particle separation), we may choose $\sigma$ so that the
distribution of the rescaled velocity differences at $\,L=\infty$
tends to the Kraichnan or to the frozen model ones
(with the notable exception of the $\,\mu\to0\,$ limit in domain A).
\vskip 0.5cm

Consider first $\,\mu\to\infty$ at $\,L=\infty$. \,Taking
$\,\sigma={1\over2(1-\alpha-\beta)}$ fixes the ratio
$\,{2D_{2,\mu}\over D_{3,\mu}}\,$ with $\,D_{i,\mu}\to\infty\,$ if
$\,\alpha+\beta>1\,$  (domain A in Fig.~\ref{fig2}) or if
$\,\alpha+2\beta<1$ (domain C in Fig.~\ref{fig2}). The latter case leads
to a non-singular Kraichnan ensemble of velocity differences with
$\,\xi=2(\alpha+\beta)\,$ whereas the former one does not (it
would correspond to $\,\xi>2,\ L=\infty$). \,We may then expect 
that \qq
{\hspace{-0.32cm}\lim\limits_{\mu\to\infty}\ \mu^\sigma\,
\CP(\mu^\sigma\rho_0,\mu^\sigma \rho;\mu t)\ \ \,=\ \CP^{Kr}
(\rho_0,\rho;t)\atop \hspace{-0.15cm}\lim\limits_{\mu\to\infty}\
\mu\,\,\CQ(\mu t; \mu^\sigma\rho_0,\mu^\sigma \rho_1)\,\,\ =\
\CQ^{Kr}(t;\rho_0,\rho_1)}\ \bigg\}\quad{\rm for} \
\,\sigma={_1\over^{2(1-\alpha-\beta)}}\ \,{\rm and}\,\
\alpha+2\beta<1\, \label{lt1} \qqq where $\,\CP^{Kr}\,$ and
$\,\CQ^{Kr}\,$ pertain to the Kraichnan model with
$\,\xi=2(\alpha+\beta)\,$. \,This is indeed consistent with the
scaling properties of the Kraichnan model dispersion. \vskip 0.5cm

Taking $\,\sigma={1\over1-\alpha}$ fixes $\,D_{2,\mu}$ with
$\,D_{3,\mu}\to0\,$ if $\,\alpha+2\beta>1\,$ (domains A and B in
Fig.~\ref{fig2}). \,We then expect that \qq
{\hspace{-0.75cm}\lim\limits_{\mu\to\infty}\
\mu^\sigma\,\CP(\mu^\sigma \rho_0,\mu^\sigma\rho;\mu t)\ \ \,=\
\CP^{fr}(\rho_0,\rho;t)\atop
\hspace{-0.55cm}\lim\limits_{\mu\to\infty}\ \mu\,\,\CQ(\mu
t;\mu^\sigma \rho_0,\mu^\sigma\rho_1)\,\,\ =\ \CQ^{\rm
fr}(t;\rho_0,\rho_1)}\ \bigg\} \quad{\rm for}\,\ \sigma
={_1\over^{1-\alpha}}\,\ {\rm and}\,\ \alpha+2\beta>1\,,
\label{lt2} \qqq where $\,\CP^{fr}$ and $\,\CQ^{fr}$ stand
for the PDF's of the frozen velocity model with H\"older exponent
$\alpha$. Note again the consistency with the scaling properties
of the Lagrangian dispersion in the frozen model. \vskip 0.5cm

Inquiring about the short-time asymptotics of the trajectory
dispersion reverses the asymptotics. We should then have \qq
{\hspace{-1.19cm}\lim\limits_{\mu\to0}\ \mu^\sigma\,\CP(\mu^\sigma
\rho_0,\mu^\sigma \rho;\mu t)\ \ \,=\ \CP^{\rm
fr}(\rho_0,\rho;t)\atop \hspace{-0.99cm}\lim\limits_{\mu\to0}\
\mu\,\,\CQ(\mu t;\mu^\sigma \rho_0,\mu^\sigma\rho_1)\,\,\ =\
\CQ^{fr}(t;\rho_0,\rho_1)}\ \bigg\} \quad{\rm for} \
\,\sigma={_1\over^{1-\alpha}}\,\ {\rm and}\,\ \alpha+2\beta<1
\label{st1} \qqq (i.e.\,\,in domain C in Fig.~\ref{fig2}) with the same
value of the H\"{o}lder exponent $\,\alpha$, and \qq
{\hspace{-0.19cm}\lim\limits_{\mu\to0}\ \mu^\sigma\,\CP(\mu^\sigma
\rho_0,\mu^\sigma \rho;\mu t)\ =\,\CP^{Kr}(\rho_0,\rho;t)\atop
\lim\limits_{\mu\to0}\ \mu\,\,\CQ(\mu t;\mu^\sigma
\rho_0,\mu^\sigma\rho_1)\,=\,\CQ^{Kr}(t;\rho_0,\rho_1)} \
\bigg\}\quad{\rm for}\,\ \sigma={_1\over^{2(1-\alpha-\beta)}}\,\
{\rm and}\ \cases{ \alpha+\beta<1\,,\cr\alpha+2\beta>1}
\label{st2} \qqq (i.e.\,\,in domain B in Fig.~\ref{fig2}) with
$\xi=2(\alpha+\beta)\,$ for the Kraichnan model. Again, this is
consistent with the scaling of the limiting PDF's.  In summary,
the scale invariance of the statistics of the pair dispersion and
of the exit time, although broken away from the
$\,\alpha+2\beta=1\,$ line, should be restored at long and short
times. \vskip 0.4cm

It has to be stressed that the relations (\ref{lt1}) to
(\ref{st2}) are conjectural. The PDF's $\,\CP\,$ and $\,\CQ\,$ are
complicated nonlinear functionals of the quasi-Lagrangian velocity
statistics and the conjectured relations assume their continuity
in an appropriate topology, which is not obvious. In particular,
since the convergence of the rescaled velocity covariances to the
one of the Kraichnan model is very slow at long distances, there
is a potential threat for the corresponding convergence of the
rescaled PDF's $\,\CP(\rho_0,\rho;t)\,$ and
$\,\CQ(t;\rho_0,\rho_1)\,$ with $\,\rho_1<\rho_0\,$ coming
from the contribution of trajectories that venture far apart,
if such contributions are important.  Similarly,
the slow convergence to the frozen model at short distances could
create problems for the corresponding convergence of the rescaled
pair dispersion and exit time PDF's, for the latter if
$\rho_1>\rho_0$. \,Whether such effects invalidate some of the
conclusions (\ref{lt1}) to (\ref{st2}) could be, in principle,
studied in perturbation theory around the Kraichnan or frozen
model. The question of the convergence of rescaled dispersion
PDF to the Kraichan model has been recently rigorously studied
in \cite{F2,F3}. We shall further discuss the non-uniformity
of the convergence of the exit time PDF to that of the frozen model
in Sect.\,\,\ref{sec:LTS}.
\vskip 0.5cm

An important question concerns the {\bf phase diagram} of the
Lagrangian flow for the model (\ref{vc}) of quasi-Lagrangian
velocities. As mentioned above, domain A is expected to correspond to
the deterministic Lagrangian flow. The rate of separation of close
trajectories in this domain (Lyapunov or Richardson flow?
locally separating or trapping? locally recurrent or transient?) is
also an open problem. Inside domains B and C, we may try to
use the conjectured convergence (\ref{lt1}) to (\ref{st2}) of the
rescaled pair dispersion and time exit PDF's to characterize the
nature of the Lagrangian flow. This will require even more care
since some uniformity of the limits will be needed.
\vskip 0.4cm

First, we may argue that, inside domains B and C, weak compressibility
$\,\wp<{d\over4(\alpha+\beta)^2}\,$ implies that the Lagrangian
flow is stochastic. The argument assumes that the dichotomy
``deterministic versus stochastic'' may be still characterized as
in Sect.\,\ref{sec:LKR}, i.e.\,\,by the behavior of
$\,\CP(\rho_0,\rho;t)\,$ in the limit $\,\rho_0\to0$. It goes as
follows. Suppose that the relation (\ref{deter}) holds at some
point inside B or C (for all $\,t$). Then, obviously, also \qq
\lim\limits_{\rho_0\to0}\,\,\mu^\sigma\,\CP(\mu^\sigma\rho_0,
\mu^\sigma\rho;\mu t)\ =\ \delta(\rho)\label{tai} \qqq for all
$\,\mu$. \,We expect the convergence (\ref{lt1}) in domain C and
(\ref{st2}) in domain B, both resulting in the Kraichnan model
PDF's, to be uniform at short distances and hence to commute with
the $\,\rho_0\to0\,$ limit. We may then infer from
Eq.\,\,(\ref{tai}) that (\ref{deter}) holds also for the limiting
Kraichnan PDF so that $\,\wp\geq d/\xi^2\,$ where
$\,\xi=2(\alpha+\beta)$, which implies the assertion. 
\vskip 0.4cm

The analogous argument may be applied when \qq
\lim\limits_{\rho_0\to0}\,\,\int\CQ(t;\rho_0,\gamma\rho_0)\,dt\ =\
\lim\limits_{\rho_0\to0}\,\,\int\,\mu\,\CQ(\mu
t;\mu^\sigma\rho_0,\mu^\sigma\gamma\rho_0)\,dt\ =\ 1\qqq leading
to the predictions that inside the domains B and C, the Lagrangian
flow is locally trapping if $\,\wp\geq{d\over4(\alpha+\beta)^2}\,$
and locally transient if
$\,\wp<{d-2\over4(\alpha+\beta)}+{1\over2}$. The first claim
seems somewhat more trustable since it does not involve
large separations where the convergence to the Kraichnan ensemble
is slowed down. Both require additionally that no mass in the exit
time PDF escapes to infinity during the $\,\mu$-limits (\ref{lt1})
and (\ref{st2}). This should not pose a problem since
convergence to the Kraichnan model becomes very fast at long times. 
\vskip 0.4cm

In the stochastic regime, the convergence (\ref{lt1}) and
(\ref{st2}) should imply for the (positive) pair dispersion
moments the behavior \qq \langle\,\rho^n(t)\,\rangle\ =\
\CO(t^{n\over 2(1-\alpha-\beta)}) \qqq for long times in domain C
and for $\,\rho_0\to0\,$ and short times in domain B, in agreement
with the naive mean-field results (\ref{ltm2}) and (\ref{stm2}).
As for the behavior of the moments of the exit time through
$\,\gamma\rho_0\,$, the relations (\ref{st2}) and (\ref{lt1})
should imply that \qq \langle\, t^n\,1_{\{t<\infty\}}\,\rangle\ =\
\CO(\rho_0^{2n (1-\alpha-\beta)}) \qqq in domain B for small
$\,\rho_0\,$ and in domain C for large ones, irrespectively of
the character of the Lagrangian flow.
Similarly, we should obtain in domain B the convergence
\qq
\lim\limits_{\mu\to0}\,\,\int f(t)\,\,\CQ(t;\mu^\sigma\rho_0,\mu^\sigma
\gamma\rho_0)\,dt\ =\ f(0)\int\CQ^{Kr}(t;\rho_0,\rho_1)
\,dt\label{RKr}\qqq
characteristic of the Richardson flow in our terminology.
\vskip 0.5cm

Similar use of the conjectured convergences (\ref{lt2}) and
(\ref{st1}) to the frozen model PDF's in order to argue about the
Lagrangian flow dichotomies and the evolution of dispersion and
exit times moments poses two problems. First is the poor
knowledge of the flow behavior in the frozen model, see, however,
Sect.\,\ref{sec:FR1D}. Second, even more serious one, is the
non-uniform nature of the convergence that becomes slow at small
distances and long times. It is still plausible, however, that
the convergence (\ref{lt2}) is uniform enough as to imply
that \qq
\langle\,\rho^n(t)\,\rangle\ =\ \CO(t^{n\over (1-\alpha)}) \qqq
for long times and for sufficiently high $\,n\,$ in the stochastic 
regime in domains A and B.
\,Similarly, relations (\ref{lt2}) and (\ref{st1}) may still imply
that for sufficiently negative moments,\qq \langle\, t^n\,\rangle\
=\ \CO(\rho_0^{n (1-\alpha)})\label{frtc}\qqq in domains A and B for large
$\,\rho_0\,$ and in domain C for small ones, see Sect.\,\,\ref{sec:LTS}. 
The same way, the convergence (\ref{RKr}) should still hold in domain C
for test functions $\,f\,$ decaying fast at infinity and
$\,Q^{Kr}\,$ replaced by $\,\CQ^{fr}$.\vskip 0.5cm

Summarizing, we predict, with various level of confidence, that 
the Lagrangian flow is 
\vskip 0.5cm

\ \hbox to 0.9cm{\hfill}deterministic 

\ \hbox to 0.9cm{\hfill}in domain A 
\vskip 0.5cm

\ \hbox to 0.9cm{\hfill}stochastic, Richardson, locally transient

\ \hbox to 0.9cm{\hfill}in domains B and C \ for 
$\,\wp<{d-2\over^{4(\alpha+\beta)}}+{1\over 2}\,$,
\vskip 0.5cm

\ \hbox to 0.9cm{\hfill}stochastic, Richardson

\ \hbox to 0.9cm{\hfill}in domains B and C \ for 
$\,{d-2\over^{4(\alpha+\beta)}}
+{1\over 2}\leq\wp<{d\over4(\alpha+\beta)^2}\,$,
\vskip 0.5cm

\ \hbox to 0.9cm{\hfill}Richardson, locally trapping

\ \hbox to 0.9cm{\hfill}in domains B and C \ for $\,\wp\geq 
{d\over4(\alpha+\beta)^2}\,$.
\vskip 0.5cm

\noindent The degree of confidence of the predictions depends 
on which of relations (\ref{lt1}) to (\ref{st2}) was used in the
argument and with what uniformity assumptions.
The predictions are consistent with the intuition that 
increase of the compressibility degree $\,\wp\,$ enhances the trapping 
of the fluid particles. It will be interesting to confirm (or infirm)
them by further analytic arguments and by numerical simulations.
Note that, in particular, we expect that in domains B and C the
incompressible Lagrangian flow is stochastic, Richardson 
and locally transient, and that, if the dimension $\,d\geq4$, \,it
stays such, whatever compressibility.

\vskip 0.5cm

\nsection{One-dimensional frozen ensemble} \label{sec:FR1D}

\noindent The frozen model was left out from the discussion of the 
phase diagram in the last section. Our arguments were based mainly on 
the convergence of the rescaled velocity ensemble (\ref{vc}) to the 
Kraichnan model and such convergence is, of course, absent for the 
frozen model. One may expect appearance of discontinuity in the 
character of the Lagrangian flow in the frozen model limit $\,D_3\to0\,$ 
which is is very non-uniform at short distances and long times, leading 
to a strong enhancement of trapping. This effect will be analyzed 
in next Section. Here we shall try to find out what happens in the
frozen velocities which, in general, are the hard to analyze. One 
case where some analytic results may be obtained is the one-dimensional 
Gaussian model with the H\"older exponent $\alpha={1\over2}$ and 2-point 
function \qq \langle v(x)\,v(y)\rangle\ =\ \int{\ee^{ik(x-y)}\over
k^2+L^{-2}}\,{dk\over2\pi}\ =\ {_1\over^2}{L}\,\ee^{-|x-y|/L}\,,
\label{1/2} \qqq where, for simplicity, we have set $\,D_2=1\,$
(what may be always achieved by rescaling $\,v\mapsto\sqrt{D_2}\,v$). 
A simplifying feature of this case, studied already in \cite{EV}, 
is that $v(x)$ forms a stationary Markov (Ornstein-Uhlenbeck) process 
with the generator  \qq \CL\ =\ -{_{1}\over^2}{_{d^2}\over^{dv^2}}
+{_v\over^L}{_d\over^{dv}} \qqq
corresponding to the Focker-Planck harmonic oscillator Hamiltonian
\qq \CH\ =\ \ee^{-{v^2\over 2L}}\CL\,\,\ee^{\,v^2\over 2L}\ =\
-{_{1}\over^2}{_{d^2}\over^{dv^2}}+{_{v^2}\over^{2L^2}}
-{_1\over^{2L}}\,. \qqq The velocity $\,v(x)\,$ with fixed $\,x\,$
is distributed according to the invariant measure of the process
\qq d\nu(v)\ =\ {_1\over^{\sqrt{\pi L}}}\,\,\ee^{-{v^2\over
L}}\,dv\,. \label{invm} \qqq The transition probabilities of the
process are \qq p(v_0,v;t)\,dv\,&=&\,\ee^{-t\,\CL}(v_0,v)\,dv\ =\
{_1\over^{\sqrt{\pi L(1-\ee^{-2t/L})}}}\,\,\ee^{-{
(\ee^{-t/L}v_{_0}-v)^2\over L(1-\ee^{-2t/L})}}\,dv\,. \label{trpr}
\qqq In the limit $L\to\infty$, the velocity difference $\,\Delta
v(x)= v(x)-v(0)\,$ becomes the one-dimensional two-sided Brownian motion
$\,w(x)$. \,The quasi-Lagrangian Eq.\,\,(\ref{rho1}) for 
the trajectory separation
takes then the form of the steepest descent equation \qq
{d\rho\over dt}\ =\ \,w(\rho)\ =\ -{d\over d\rho} \phi(\rho)\,.
\label{stdes} \qqq in the potential $\,\phi(x)=-\,
\int\limits_0^{x}w(y)\,dy$. The solution $\,\rho(t)\,$ slides to
the bottom of the potential well in which the initial point
$\,\rho_0=\rho(0)\,$ is situated, i.e. to the closest zero
$\,\rho_+$ of $\,w(\rho)\,$ to the right of $\,\rho_0$ if
$\,w(\rho_0)>0\,$ or $\,\rho_-$ to the left of $\,\rho_0$ is
$\,w(\rho_0)<0$. \,The only difference with the case of smooth
potential with wells approximately quadratic around typical
minima, is that, as we show below, the solution will arrive to the
bottom of the well in a finite rather than infinite time. This is
due to the roughness of the Brownian motion. After arriving at the
bottom, the solution will stay locked there in subsequent
times. \vskip 0.5cm

We may restrict ourselves to the case $\,\rho_0>0$.
Suppose also that $\,w(\rho_0)>0$. The first value $\,\rho_+$
to the right of $\,\rho_0$ such that $\,w(\rho_+)=0$ is finite
with probability one. The time $\,t_+$ that the solution
$\,\rho(t)\,$ of Eq.\,\,(\ref{stdes}) starting at $\,\rho_0$ at
time zero takes to reach $\,\rho_+$ is \qq t_+\ =\
\int\limits_{\rho_0}^{\rho_+}{d\rho\over w(\rho)}\,. \label{time1}
\qqq Let us compute the expectation
$\,\langle\,t_+\,1_{\{w(\rho_0)>0,\, \rho_+\leq\rho_2\}}\rangle\,$
of times $\,t_+$ over the Brownian paths $\,w(\rho)\,$ such that
$\,w(\rho_0)>0\,$ and $\,\rho_+\leq \rho_2\,$ for certain
$\,\rho_2>\rho_0$. It is equal to \qq
\int\limits_{\rho_0}^{\rho_2}\hspace{-0.06cm}d\rho\hspace{-0.06cm}
\int\limits_0^{\infty}\hspace{-0.06cm}\ee^{-{w_0^2\over2\rho_0}}
{_{dw_0}\over^{\sqrt{2\pi\rho_0}}}\hspace{-0.06cm}
\int\limits_0^\infty\hspace{-0.06cm}\Big(\ee^{-{(w_0-w)^2 \over
2(\rho-\rho_0)}}-\,\ee^{-{(w_0+w)^2\over 2(\rho-\rho_0)}}\Big)
{_{dw}\over^{\sqrt{2\pi(\rho-\rho_0)}\,w}}
\hspace{-0.01cm}\int\limits_{0}^{\infty}\hspace{-0.06cm}
\ee^{-{(w+w_2)^2\over2(\rho_2-\rho)}}{_{2\,dw_2}
\over^{\sqrt{2\pi(\rho_2-\rho)}}}\,. \label{avt1+} \qqq The origin
of this formula is straightforward. The probability that
$\,w(\rho_0)\,$ belongs to $\,[w_0,w_0+dw_0]\,$ is $\,
\ee^{-{w_0^2\over2\rho_0}}{dw_0\over\sqrt{2\pi\rho_0}}$. \,The one
that $\,w(\rho)\,$ belongs to $\,[w,w+dw]\,$ without passing
through zero between $\,\rho_0\,$ and $\,\rho\,$ is
$\,\Big(\ee^{-{(w_0-w)^2\over 2(\rho-\rho_0)}}
-\,\ee^{-{(w_0+w)^2\over 2(\rho-\rho_0)}}\Big){dw\over\sqrt{2\pi
(\rho-\rho_0)}}$, given that $\,w(\rho_0)=w_0$, (it is expressed
by the heat kernel with the Dirichlet condition at $\,w=0$).
\,Finally, the last integral on the right hand
side of (\ref{avt1+}) is equal to the probability that the Brownian 
trajectory crosses zero between $\,\rho\,$ and $\,\rho_2$,
given that $\,w(\rho)=w$. \,In Appendix\,\,1 we show that \qq
\langle\,t_+\,1_{\{w(\rho_0)>0,\,\rho_+\leq\rho_2\}}\rangle\ \leq\
{C_+}\,\rho_0^{1/2}\,\ln{_{\rho_2}\over^{\rho_0}}\,, \label{posa}
\qqq where $\,C_+\,$ is a dimensionless constant. This proves that
the time $\,t_+\,$ is almost surely finite although the
unrestricted mean $\,\langle\,t_+\,1_{\{w(\rho_0)>0\}}\,\rangle$,
\,given by the $\,\rho_2\to\infty\,$ limit of (\ref{avt1+}) under
which the last integral on the right hand side tends to one,
diverges. The divergence is due to the contribution of the
Brownian paths that travel far before falling back to zero. 
In the Ornstein-Uhlenbeck process with $\,L<\infty$, \,the weight 
of such paths is suppressed and it is not difficult to show that 
$\,\langle\,t_+\,1_{\{w(\rho_0)>0\}}
\,\rangle\,$ is finite then. \vskip 0.4cm

Similarly, let $\,w(\rho_0)<0$ and $\,0\leq\rho_-<\rho_0$ be the
first value to the left of $\,\rho_0$ such that $\,w(\rho_-)=0$.
The time $\,t_-$ that the solution $\,\rho(t)\,$ of
Eq.\,\,(\ref{stdes}) starting at $\,\rho_0$ at time zero takes to
reach $\,\rho_-$ is given by Eq.\,\,(\ref{time1}) with $\,t_+$
replaced by $\,t_-$ and $\,\rho_+$ by $\,\rho_-$. \,The
expectation value $\,\langle\,t_-\,1_{\{w(\rho_0)<0\}}\,\rangle\,$
is given by \qq -\int\limits_0^{\rho_0}d\rho
\int\limits_{-\infty}^0\ee^{-{w^2\over2\rho}}
{_{dw}\over^{\sqrt{2\pi\rho}\,\,w}}
\int\limits_{-\infty}^0\Big(\ee^{-{(w-w_0)^2\over 2(\rho_0
-\rho)}}-\,\ee^{-{(w+w_0)^2\over 2(\rho_0-\rho)}}\Big)
{_{dw_0}\over^{\sqrt{2\pi(\rho_0-\rho)}}} \label{avt1-} \qqq which
is easily seen to be finite, e.g.\,\,by bounding the last integral
by $\,\sqrt{2\over\pi(\rho_0-\rho)}\,|w|$. We infer that \qq
\langle\,t_-\,1_{\{w(\rho_0)<0\}}\,\rangle\ =\ {C_-}
\,\rho_0^{1/2}\,, \qqq where $\,C_-$ is another
dimensionless constant. \vskip 0.5cm

\subsection{\ Exit time}

\noindent As for the exit time that the process $\,\rho(t)\,$ takes 
to grow from $\,\rho_0\,$ to $\,\rho_1>\rho_0$, \,it is finite if and 
only if $\,\rho(t)\,$ is not stuck at a zero $\,\rho_\pm<\rho_1$ of
$\,w(\rho)$, i.e.\,\,if $\,w\,$ is positive on the interval
$\,[\rho_0,\rho_1)$. \,It is then given by the formula \qq t\ =\
\int\limits_{\rho_0}^{\rho_1}{d\rho\over w(\rho)}\,. \label{time2}
\qqq The probability that $\,t<\infty\,$ is given by \qq
\langle\,1_{\{t<\infty\}}\,\rangle\ =\ \int\limits_0^\infty
\ee^{-{w_0^2\over2\rho_0}}{_{dw_0}\over^{\sqrt{2\pi\rho_0}}}
\int\limits_{0}^{\infty}\Big(\ee^{-{(w_0-w_1)^2\over2(\rho_1-\rho_0)}}
-\,\ee^{-{(w_0+w_1)^2\over2(\rho_1-\rho_0)}}\Big)
{_{dw_1}\over^{\sqrt{2\pi(\rho_1-\rho_0)}}}\,. \label{11} \qqq
Note that the last expression depends only on $\,{\rho_1\over\rho_0}$,
\,is smaller than $\,1/2\,$ and tends to zero when $\,\rho_0\to0\,$
with $\,\rho_1$ fixed. \,With the complementary probability, the
solution $\,\rho(t)\,$ starting from $\,\rho_0>0\,$ will never
reach $\,\rho_1$, \,i.e.\,\,the exit time $\,t\,$ is infinite.
\,The averages of positive powers of the exit time constraint to be finite
are expressed by the relation \qq
\langle\,t^n\,1_{\{t<\infty\}}\,\rangle\ =\
n!\hspace{-0.6cm}\int\limits_{\rho_0\leq\rho'\leq\dots\leq\rho^{(n)}
\leq\rho_1}\hspace{-0.6cm}d\rho'\cdots d\rho^{(n)}\
\int\limits_0^{\infty}
\ee^{-{w_0^2\over2\rho_0}}{_{dw_0}\over^{\sqrt{2\pi\rho_0}}}
\hspace{2cm}\,\,\cr \cdot\ \prod\limits_{i=1}^n
\int\limits_0^\infty \Big(\ee^{-{(w^{(i-1)}-w^{(i)})^2} \over
2(\rho^{(i)}-\rho^{(i-1)})}-\,\ee^{-{(w^{(i-1)}+w^{(i)})^2}\over
2(\rho^{(i)}
-\rho^{(i-1)})}\Big){_{dw^{(i)}}\over^{\sqrt{2\pi(\rho^{(i)}-\rho^{(i-1)})}
\,\,w^{(i)}}}\,\,\,\cr \cdot\ \int\limits_0^\infty
\Big(\ee^{-{(w^{(n)}-w_1)^2\over
2(\rho_1-\rho^{(n)})}}-\,\ee^{-{(w^{(n)}+w_1)^2\over
2(\rho_1-\rho^{(n)})}}\Big){_{dw_1}\over^{\sqrt{2\pi(\rho_1-\rho^{(n)})}}}
\label{avtn} \qqq with $\,\rho^{(0)}\equiv\rho_0\,$ and
$\,w^{(0)}\equiv w_0$. \,It is easy to show that the expression on
the right hand side is finite. Indeed, bounding the last integral
by $\,\sqrt{2\over\pi(\rho_1- \rho^{(n)})}w^{(n)}\,$ as in
estimating (\ref{avt1-}) and proceeding further the same way, we
obtain the inequality \qq \langle\,t^n\,1_{\{t<\infty\}}\,\rangle\
\leq\ {_1\over^\pi}\left
({_2\over^{\pi}}\right)^{n/2}\sqrt{\rho_0}\,\
n!\hspace{-0.5cm}\int\limits_{\rho_0\leq\rho'\leq\dots\leq\rho^{(n)}
\leq\rho_1}\hspace{-0.5cm}{_{d\rho'\cdots
d\rho^{(n)}}\over^{\sqrt{(\rho'-\rho_0)\cdots
(\rho_1-\rho^{(n)})}}}\,\,\cr \leq\ {_{2^n}\over^{\sqrt{2\pi}}}\,
{_{n!}\over^{(n-1)!!}}\,({_{\rho_1}\over^{\rho_0}}-1)^{^{n-1\over2}}\,
\rho_0^{^{n\over2}}\,, \label{bds} \qqq where the last line
results from the inductive calculation of the $\rho^{(i)}$
integrals. As a consequence of the scaling properties of the right
hand side of Eq.\,\,(\ref{avtn}), \qq
\langle\,t^n\,1_{\{t<\infty\}} \,\rangle\ =\ {C_n}\,\rho_0^{n/2}\ =\ 
C_n\,\tau_e^n\,
\label{above} \qqq where $\,C_n\,$ are dimensionless constants
depending on the ratio $\,\gamma={\rho_1\over\rho_0}>1\,$
and $\,\tau_e=(\rho_0/D_2)^{1/2}\,$ is the eddy turnover 
time at scale $\,\rho_0\,$ (recalled that we have set $D_2=1$ above).
\vskip 0.5cm

The bounds (\ref{bds}), which respect the above scaling, imply
that the characteristic function $\,\langle\,\ee^{i\omega
t}\,1_{\{t<\infty\}}\,\rangle\,$ is entire in $\omega$ and that
the large $\,t\,$ decay of the PDF $\,\CQ(t;\rho_0,\rho_1)\,$ is
at least $\,\propto \ee^{-\CO(t^2)}$.
\,Due to Eq.\,\,(\ref{time2}), the characteristic function 
may be expressed by the path integral  \qq \langle\,\ee^{i\omega
t}\,1_{\{t<\infty\}}\,\rangle\,\ =\ {_1\over^\CN}\,
\int\limits_0^\infty\ee^{-{w_0^2\over2\rho_0}}{_{dw_0}\over
^{\sqrt{2\pi\rho_0}}}
\int\ee^{\,\int\limits_{\rho_0}^{\rho_1}\left({i\omega\over
w(\rho)}\,-\,{1\over 2}{\dot{w}}^2(\rho)\right)d\rho}\,\CD w
\label{pathin}\qqq  over the paths
$\,[\rho_0,\rho_1]\ni\rho\mapsto w(\rho)\in(0,\infty)\,$ such that
$\,w(\rho_0)=w_0$, \,with $\,\CN\,$ being an appropriate normalization
factor. The expression permits to evaluate the large 
$\,|\omega|$-behavior along
the positive imaginary axis of $\,\omega\,$ by the semi-classical
calculation. The extremal trajectory $\,w(\rho)\,$ describes a
motion of a unit mass particle climbing up in the potential
$\,-{1\over w}\,$ until a total stop. It satisfies the equation
\qq {_{\sin{\varphi_0}}\over^{\cos^3{\varphi_0}}}\left(\varphi
+\sin{\varphi}\cos{\varphi} \right)\ =\
{_{\rho_1}\over^{\rho_0}}-{_{\rho}\over^{\rho_0}}\label{cle} \qqq
for $\,\cos^2{\varphi}={w(\rho)\over w(\rho_1)}$,
$\,\cos^2{\varphi_0}={w(\rho_0)\over w(\rho_1)}\,$ and
$\,\varphi\,$ between zero and $\,\varphi_0<{\pi\over2}$. \,In
particular, $\,\varphi_0\,$  is determined by Eq.\,\,(\ref{cle})
with $\,\varphi=\varphi_0\,$ and $\,\rho=\rho_0$. \,It depends
only on the ratio $\,{\rho_1\over\rho_0}$. \,The action of the
classical trajectory is $\,S_0=|\omega|^{2/3}\rho_0^{1/3}s_0\,$ with
a dimensionless constant $\,s_0={3\varphi_0\,\sin^{1/3}{\varphi_0}
\over2^{1/3}\cos{\varphi_0}}\,$ growing with the
$\,{\rho_1\over\rho_0}$. \,The decay $\,\sim\ee^{-S_0}\,$ of the
characteristic function along the positive imaginary axis
corresponds to the small $\,t\,$ behavior
$\,\sim\ee^{-{4\over27}\,\rho_0\,s_0^3\,t^{-2}}=
\ee^{-{4\over27}\,s_0^3\,(\tau_e/t)^{2}}\,$ of the exit
time PDF $\,\CQ(t;\rho_0,\rho_1)\,$ (up to powers of
$\,t$). \,One could expect that for short times the
trajectories move almost ballistically: \qq \rho(t)\ \simeq\
\rho_0+w(\rho_0)t\qqq which would give $\,t\simeq{\rho_1-\rho_0\over
w(\rho_0)}\,$ and a short time tale
$\,\sim\ee^{-{(\rho_1-\rho_0)^2\over2\rho_0}\,t^{-2}}\,$ of the
exit time PDF. This reproduces well the power of $\,t\,$ in the
exponential but not the coefficient. The latter, divided by
$\,\rho_0$, \,agrees only to the order
$\,({\rho_1\over\rho_0}-1)^2$ with the correct expression
$\,{4\over 27}s_0^3$. \vskip 0.5cm

For $\,\omega\,$ on the imaginary axis, the path-integral on the
right hand side of Eq.\,\,(\ref{pathin}) may be re-expressed in the
operator language via the Feynman-Kac formula. Using also the
invariance of the Brownian motion under the scale transformations
$\,w(x)\mapsto|\omega|^{-1}w(|\omega|^2x)$, \,we obtain the
identity \qq \langle\,\ee^{i\omega
t}\,1_{\{t<\infty\}}\,\rangle\,\ =\
\int\limits_0^\infty\ee^{-{w_0^2\over2|\omega|^2\rho_0}}{_{dw_0}\over
^{\sqrt{2\pi\rho_0}\,|\omega|}} \int\limits_0^\infty
\ee^{-|\omega|^2(\rho_1-\rho_0)\CK_{_\pm}}(w_0,w_1) \,dw_1\,,
\label{eck} \qqq where the operator \qq \CK_\pm\ =\
-{_1\over^2}{_{d^2}\over^{dw^2}}\pm{_{1}\over^{w}} \label{ck} \qqq
on the interval $\,[0,\infty)\,$ is a 1-dimensional Schr\"{o}dinger 
operator with Dirichlet boundary condition at zero. The
signs pertain to the positive or negative imaginary
$\,\omega$-axis and result in a repulsive or an attractive
potential, respectively. An explicit expression for the kernel of
the exponential of $\,\CK_-$, \,see Appendix\,\,2, shows the
growth of the characteristic function for
$\,\omega=-i|\omega|\,$ dominated by the lowest bound state
contribution $\,\sim\ee^{\,-|\omega|^2(\rho_1-\rho_0)E_0}\,$ with
$\,E_0=-{1\over2}$. \,This implies the decay
$\,\sim\ee^{-{1\over(\rho_1 -\rho_0)}\,t^2}
=\ee^{-{1\over\gamma-1}\,(t/\tau_e(\rho_0))^2}\,$ of the PDF
$\,\CQ(t;\rho_0,\rho_1)\,$ for large $\,t$.
\vskip 0.4cm

\begin{figure}
\begin{center}
\mbox{\hspace{0.0cm}\epsfig{file=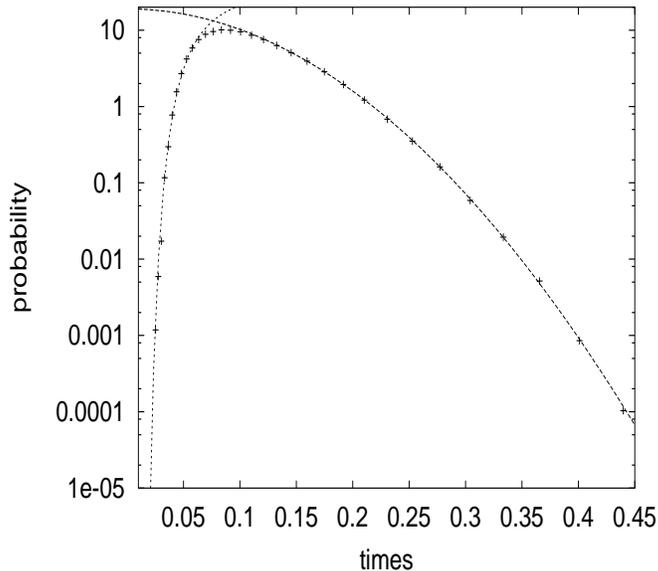,height=8cm,width=9cm}}
\end{center}
\vspace{0.3cm}
\caption{The PDF of the exit times for the frozen flow with $\alpha=1/2$.
The dotted curves are the analytical predictions at small and large
times, respectively.}
\label{fig:pdf_0.5}
\end{figure}

The above analytic predictions may be used to validate the
numerical method to be applied later for the cases $\alpha\neq 1/2$,
where such rigorous results are not available. The velocity field is
generated by using a straightforward Fourier method, i.e. generating
the Fourier modes by a standard Gaussian random number generator and
transforming back to real space by Fast Fourier Transform.  The
resolution which we used was $2^{20}$. The initial separation is set at
$20,000$ and we measure the PDF of the time taken to reach $5$ times 
the initial separation. The curve is shown in Fig.~\ref{fig:pdf_0.5}.  
The agreement with the superposed predictions indicates that the 
choice of the resolution and the initial separation are appropriate 
to avoid contamination by periodicity and/or discretization effects.
\vskip 0.4cm

\vskip 0.5cm
The time in which $\,\rho(t)\,$ decreases from $\,\rho_0\,$ to a
positive value $\,\rho_1<\rho_0\,$ is finite and still determined by
Eq.\,\,(\ref{time2}) if and only if
$\,w<0\,$ on the interval $\,(\rho_1,\rho_0]$.
\,It is easy to see that
\qq
\CQ(t;\rho_0,\rho_1)\ =\ \CQ(t;\rho_1,\rho_0)
\label{sy}
\qqq
since for $\,\rho_1<\rho_0\,$ the characteristic function
$\,\langle\,\ee^{i\omega t}\,1_{\{t<\infty\}}\,\rangle\,$ is given
by the expression (\ref{pathin}) with $\,\rho_0\,$ and
$\,\rho_1\,$ interchanged.
\vskip 0.5cm

The scaling property (\ref{scp}) with $\,\sigma=2\,$ and the
fact that the exit time is infinite with a positive probability
depending on $\,{\rho_1\over\rho_0}\,$ imply that
\qq
\lim\limits_{\rho_0\to0}\,\, \CQ
(t;\rho_0,\gamma\rho_0)=c(\gamma)\,\delta(t)
\label{ntnr}
\qqq
for any positive $\,\gamma\not=1\,$ with $\,0<c(\gamma)<1$.
\,It follows that the Lagrangian flow is Richardson. locally trapping 
and locally transient in the terminology of Sect.\,\ref{sec:LKR}A.
\vskip 0.5cm


\subsection{\ Pair dispersion}

\noindent The pair dispersion in the model is closely related to 
the exit time. The reason is that the solution $\,\rho(t)\,$ 
of Eq.\,\,(\ref{stdes}) never changes direction. As a result, 
the solution that starts at $\,\rho_0\,$ at time zero satisfies
$\,\rho(t)\geq\rho_1\geq\rho_0\,$ or
$\,\rho(t)\leq\rho_1\leq\rho_0\,$ if and only if it exits through
$\,\rho_1\,$ in time shorter than $\,t$. \,It follows that \qq &&
\int\limits_{\rho_1}^\infty \CP(\rho_0,\rho;t)\,d\rho\ =\
\int\limits_0^t\CQ(s;\rho_0,\rho_1)\,ds\qquad{\rm for}\quad
\rho_1\geq\rho_0\,,\label{trh1}\\&& \int\limits_0^{\rho_1}
\CP(\rho_0,\rho;t)\,d\rho\ =\
\int\limits_0^t\CQ(s;\rho_0,\rho_1)\,ds\qquad{\rm for}\quad
\rho_1\leq\rho_0\,\,\label{trh2}\qqq or that \qq
\CP(\rho_0,\rho;t)\ =\ \mp\,\partial_{\rho}\int\limits_0^t
\CQ(s;\rho_0,\rho)\,ds\qquad{\rm
for}\quad\rho\,\,{_{_>}\atop^{^<}} \,\m\rho_0\,.\label{PQ} \qqq
The right hand side of Eq.\,\,(\ref{trh1}) is bounded by
$\,\int\limits_0^\infty\CQ(s;\rho_0,\rho_1)\,ds\,$ which tends to
zero when $\,\rho_0\to0\,$ for $\,\rho_1>0\,$ fixed, \,see
Eq.\,\,(\ref{11}). It follows that \qq \lim\limits_{\rho_0\to0}\
\CP(\rho_0,\rho;t)\ =\ \delta(\rho)\,, \qqq in agreement with the
fact that the trajectories are determined by the initial
condition. Hence the Lagrangian flow is deterministic. Note the 
important difference with the phase {\bf V} of the Kraichnan
model, see Fig.~\ref{fig1}. There the trapping effects 
resulted in the coalescence 
of fluid particles signaled by the singular form (\ref{scwd}) 
of the pair dispersion
PDF. With the use of symmetry (\ref{sy}), the left hand side of
Eq.\,\,(\ref{trh2}) may be bounded by $\,\int\limits_0^\infty
\CQ(s;\rho_1,\rho_0)\,ds\,$ and hence tends to zero when
$\,\rho_1\to0\,$ for fixed $\,\rho_0>0$. \,As a result, the PDF
$\,\CP(\rho_0,\rho;t)\,$ cannot have a contribution proportional
to $\,\delta(\rho)\,$ for $\,\rho_0>0$. \,It is, instead, regular 
in $\,\rho$. \, The reason is that in the quasi-Lagrangian model of
particle separation (\ref{stdes}), the solutions $\,\rho(t)\,$ are 
trapped with probability 1 at positive zeros of $\,w\,$ (there is an
infinity of such zeros) and never arrive at $\,\rho=0$. This seems
to be an artifact of the frozen one-dimensional model but it serves 
as a warning that the behavior of trajectories in the time-correlated
velocities may be richer than what was observed for the Kraichnan
model, with a possible occurrence of phase \vskip 0.5cm

\ \hbox to 0.9cm{{\bf VI}.\hfill} \ deterministic, Richardson,
locally trapping, locally transient
\vskip 0.5cm

\noindent characterized by a combination of properties that did not 
occur in the time-decorrelated model.
\vskip 0.3cm

\subsection{\ Case with $\,\alpha\not={1\over2}$}

\noindent Several of the features analytically established for the
one-dimensional frozen model with the H\"older 
exponent $\,\alpha={1\over2}\,$
may be generalized to the case of general $\,\alpha\,$ in the
interval $\,(0,1)$, \,although the absence of the Markov property in 
the process $\,v(x)\,$ makes the arguments more difficult, see \cite{HKW}. 
In the limit $\,L\to\infty$, \,the equation for the trajectory
separation still has the form (\ref{stdes}) with $\,w(x),\
x\geq0,\,$ being (upon a right choice of the normalization
constant $D_2$) the two-sided {\bf fractional Brownian motion} (fBm),
i.e.\,\,the Gaussian process with mean zero and 2-point function
\vskip 0.3cm
\qq \langle w(x)\,w(y)\rangle\ =\
{_1\over^2}(x^{2\alpha}+y^{2\alpha}- |x-y|^{2\alpha})\ \equiv\
G(x,y)\, \label{fbm} \qqq 
\vskip 0.1cm
\noindent for $\,x,\,y\geq 0$. \,The 2-point function 
(\ref{fbm}) is a
kernel of a positive operator on the half-axis $\,[0,\infty)\,$ that 
we shall denote by $\,G$. Note the scale invariance under
$\,w(x)\mapsto\mu^{-\alpha}w(\mu x)\,$ of the fBm. The basic
result of \cite{Molch} asserts that the probability that
$\,w(x)<w_0\,$ for $\,w_0>0\,$ and all $\,x\,$ in the interval
$\,(0,\rho_0)\,$ behaves like $\,\CO((\rho_0w_0^{-1/\alpha}
)^{1-\alpha})\,$ for large values of $\,\rho_0w_0^{-1/\alpha}$. \,
Similarly as for the Brownian motion, the fBm lives on continuous
trajectories and has zeros in any interval $\,(0,\rho_0)\,$ or
$\,(\rho_0,\infty)$. One may again prove that, with probability 1,
the solution $\,\rho(t)\,$ of Eq.\,\,(\ref{stdes}) starting at any
$\,\rho_0>0\,$ arrives in finite time at the closest zero $\,\rho_\pm\,$ 
to the right or left of $\,\rho_0$, \,see \cite{HKW}. 
\vskip 0.6cm

As in the case $\,\alpha={1\over2}$,
\,the exit time $\,t\,$ through $\,\rho_1>\rho_0\,$ is
finite if and only if $\,w>0\,$ on the
interval $\,[\rho_0,\rho_1)\,$ and the probability
$\,\langle\,1_{\{t<\infty\}}\,\rangle\,$ of such an event
depends only on $\,{\rho_1\over\rho_0}\,$ and tends to zero when
$\,\rho_0\to0$.
It should be again possible to extract the behaviors of the exit
time PDF $\,\CQ(t;\rho_0,\rho_1)\,$ for large and small
time by looking at the large $\,|\omega|\,$ behavior of \qq
\langle\,\ee^{\pm|\omega|t}\,1_{\{t<\infty\}}\,\rangle\ =\
\langle\,\ee^{\pm|\omega|\int\limits_{\rho_0}^{\rho_1}{d\rho\over
w(\rho)}}\,1_{\{w>0\ {\rm on}\ [\rho_0,\rho_1)\}}\,\rangle\,,
\label{fbch} \qqq 
\vskip 0.2cm
\noindent where the last expectation is with respect to
the Gaussian measure of the fBm $\,w$. \,For\nobreak\ \nobreak the 

\noindent negative sign,
this expression should be still dominated for large $\,|\omega|\,$
by the semi-classical contribution $\,\sim\ee^{-S_0}$. \,The
classical trajectory
$\,w(\rho)=|\omega|^{1/3}\rho_0^{(1+2\alpha)/3}
(Gu)({\rho\over\rho_0}),$ \,where $\,u\,$ is a function that does
not vanish only on the interval $\,(1,{\rho_1\over\rho_0})\,$ and
such that \qq u\ =\ (Gu)^{-2} \qqq there. Note that it follows
that $\,w(\rho)>0\,$ for $\,\rho>0\,$ since $\,G(x,y)>0\,$ except
for $\,x,y=0$. \,The action of the classical trajectory
is $\,S_0=|\omega|^{2/3}\,\rho_0^{2(1-\alpha)/3}\,s_0\,\,$ for \qq
s_0\ =\ \int\limits_{1}^{\rho_1/\rho_0}\Big[{_1\over^{(Gu)(x)}}
\,+\,{_1\over^2}\, u(x)\,(Gu)(x)\Big]\,dx\,. \qqq Such a
semi-classical dominance implies again the small time tail
$\,\,\sim\ee^{-{4\over27}\,\rho_0^{2(1-\alpha)}\,s_0^3\,t^{-2}}=
\ee^{-\CO((\tau_e/t)^2)}\,$
of the exit time PDF $\,\CQ(t;\rho_0,\rho_1)\,$ (with the eddy 
turnover time $\,\tau_e=D_2^{-1/2}\rho_0^{1-\alpha}$).
\vskip 0.6cm

\begin{figure}
\begin{center}
\mbox{\hspace{0.0cm}\epsfig{file=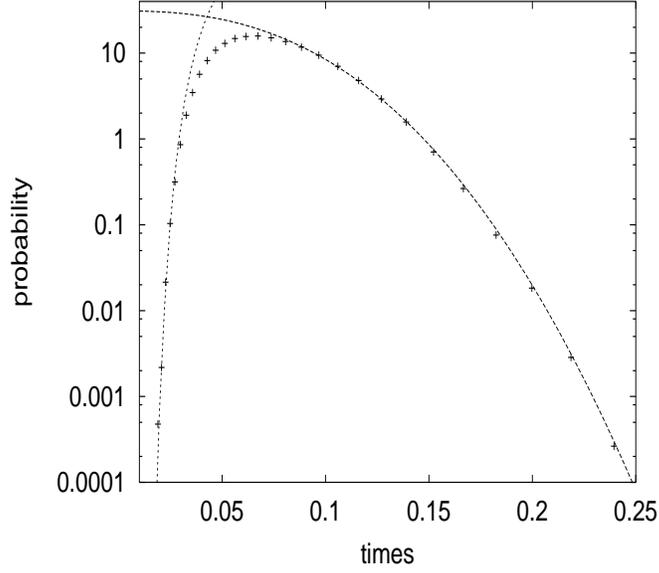,height=8cm,width=9cm}}
\end{center}
\vspace{0.3cm}
\caption{The PDF of the exit times for the 1$d$ frozen velocities with $\alpha=0.4$.
The dotted curve is a fit of the form $\ee^{-{\rm const.}\,t^{1/\alpha}}$.}
\label{fig:pdf_0.4}
\end{figure}

\begin{figure}
\begin{center}
\vspace{-0.7cm}
\mbox{\hspace{0.0cm}\epsfig{file=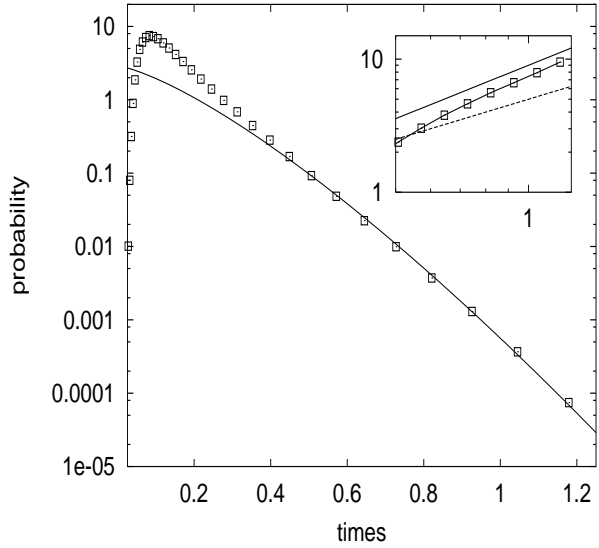,height=8cm,width=9cm}}
\end{center}
\vspace{0.3cm}
\caption{The PDF of the exit times for the 1$d$ frozen velocities with $\alpha=3/4$.
The solid curve is a fit of the form $\ee^{-{\rm
const.}\,t^{1/\alpha}}$. Since the curve gives the visual impression
of an exponential decay, we also plot in the inset
$\ln|\ln(\CQ(t;\rho_0,\rho_1))|$ {\it vs} $\log t$ and compare the
slopes $0.75$ and $1$ to validate the former.}
\label{fig:pdf_0.75}
\end{figure}

With the use of the scale invariance of the fBm, one may also absorb 
the $\,|\omega|$-dependence of the characteristic
function into the length of the $\,\rho$-interval in 
Eq.\,\,(\ref{fbch}): \qq
\langle\,\ee^{\pm|\omega|t}\,1_{\{t<\infty\}}\,\rangle\ =\
\langle\,\ee^{\pm\int\limits_{\rho'_0}^{\rho'_1}{d\rho\over
w(\rho)}}\,1_{\{w>0\ {\rm on}\ [\rho'_0,\rho'_1)\}}\,\rangle\,,
\label{fbch+} \qqq where
$\,\rho'_i=|\omega|^{1\over1-\alpha}\rho_i$. \,For the positive sign
in Eq.\,\,(\ref{fbch+}) and $\,\alpha={1\over2}$, \,we have used the
Feynman-Kac formula in order to extract the extensive behavior of the
right hand side for large $\,|\omega|$. \,For other values of
$\,\alpha\in(0,1)\,$ such a formula is not available but the
expectation (\ref{fbch+}) may be viewed as the partition function of a
one-dimensional continuous spin system with long-range 2-spin
interactions decaying as $\,distance^{-2(1+\alpha)}\,$ and with
partially confining (as opposed to the case with negative sign)
single-spin potential. It is plausible that the extensive behavior of
the partition function
$\,\,\sim\,\ee^{-|\omega|^{_1\over^{1-\alpha}}(\rho_1-\rho_0)
E_0}\,\,$ for large $\,|\omega|\,$ with free energy density
$\,E_0<0\,$ persists for such systems. Such a behavior would result in
the long-time tail
$\,\,\sim\,\ee^{-\alpha({1-\alpha\over|(\rho_1-\rho_0)E_0|})^{
(1-\alpha)/\alpha}\,t^{1/\alpha}}
=\ee^{-\CO((t/\tau_e)^{1/\alpha})}\,$ of the PDF
$\,\CQ(t;\rho_0,\rho_1)$. \,The validity of this prediction is
confirmed by Figs.~\ref{fig:pdf_0.4} and \ref{fig:pdf_0.75} where we
present the PDF's of the exit times for the two cases $\alpha=0.4$ and
$0.75$. The numerical simulations are realized by the same method
previously validated in the case $\alpha=1/2$.
\vskip 0.6cm

The statistics of the time of exit through $\,\rho_1<\rho_0\,$ is
again obtained from that for $\,\rho_1>\rho_0\,$ by interchanging
$\,\rho_0\,$ and $\,\rho_1$ and the pair dispersion PDF is still given
by Eq.\,\,(\ref{PQ}). The remarks about the phase {\bf VI} type
behavior of Lagrangian flow carry over from the case
$\,\alpha={1\over2}$.

\vskip 0.6cm

\bigskip

\nsection{Effect of long time velocity correlations}
\label{sec:LTS}

The presence of permanent trapping of trajectories in one-dimensional
frozen ensemble, leading to events with infinite exit time through
$\,\rho_1=\gamma\rho_0$, \,should also occur in higher-dimensional frozen 
ensembles obtained by setting $\,D_3=0\,$ in Eq.\,\,(\ref{vc}). It is 
plausible that, at least in the presence of compressibility,  the sets 
of velocities trapping trajectories in a given region have positive 
probability. If, however, one reintroduces finite temporal correlations 
of velocities by taking positive $\,D_3\,$ in (\ref{vc}), the particles 
will eventually be released from traps after the time evolution changes 
the velocity field configuration, i.e.\,\,after time of order 
$\,\tau_c(\rho)=D_3^{-1}\rho^{2\beta}\,$ for traps of size $\,\rho$. 
\,If for $\,D_3>0\,$ the exit times are finite so that 
$\,\int\limits_0^\infty\CQ(t)\,dt=1$, \,then the missing probability 
$\,1-\int\limits_0^\infty\CQ(t)\,dt>0\,$ for $\,D_3=0\,$ should be built 
from the long-time tails of $\,\CQ\,$ at positive $\,D_3$, \,as the 
latter is taken to zero. We may then expect that, loosely 
speaking, for fixed $\,\rho_0$, $\,\gamma\,$ and $\,D_2$, 
\qq
\CQ(t;\rho_0,\gamma\rho_0)\ \approx 
\tau_c^{-1}\,\CQ^{\infty}(t/\tau_c,\gamma)\qquad
{\rm for}\quad t/\tau_c\gg 1\,.
\label{lart}
\qqq
with $\,\tau_c=\tau_c(\rho_0)\,$ and the function $\,\CQ^{\infty}\,$ having a (Poissonian)
exponential tail. Let us concentrate on the one-dimensional situation
where the frozen exit-time PDF is expected to have a stretched exponential
tail  $\,\sim\ee^{-\CO((t/\tau_e)^{1/\alpha})}\,$ with the scale set by the eddy
turnover time $\,\tau_e=D_2^{-1/2}\rho_0^{1-\alpha}$. \,Note that
the latter attains the value of order $\,\tau_c^{-1}\,$ for 
$\,t=\tau'\,$ with $\,\tau'/\tau_e=\CO((\ln(\tau_c/\tau_e))^\alpha)
=\CO(|\ln D_3|^\alpha)$.
The ``minimal'' scenario would be that the frozen PDF passes into
the form of (\ref{lart}) around $\,t=\tau'$.   
\,More precisely, we may postulate the convergence
\qq
\hbox to 5.6cm{$\CQ(t;\rho_0,\gamma\rho_0)\,\,1_{\{t<\tau'\}}$\hfill}& 
\mathop{\longrightarrow}\limits_{D_3\to0}&\ \quad\CQ^{fr}(t;\rho_0,\gamma\rho_0)\cr
& &\label{scen}\\
\hbox to 5.6cm{$\tau_c\,\,\CQ(s\tau_c;\rho_0,
\gamma\rho_0)\,\,1_{\{s\geq\tau'/\tau_c\}}$\hfill}&
\mathop{\longrightarrow}\limits_{D_3\to0}&\ \quad\CQ^{\infty}(s;\gamma) 
\nonumber
\qqq
in a strong enough sense. The above relations imply that the large exit 
time behavior becomes self-similar for small $\,D_3$, \,with the 
characteristic scale equal to $\tau_c$, \,with no intermediate 
regime between the frozen type behavior and the self-similar tail.
It is also possible that a different intermediate regime sets in
between times of order $\,\tau'\,$ and $\,\tau_c$, with $\,\tau'\,$
depending differently on $\,D_3$.   
\vskip 0.4cm

The minimal scenario would make explicit the large $\,t\,$
non-uniformity of the conjectured convergence of the exit time PDF's,
see (\ref{lt2}) and (\ref{st1}).  Recall that the conjecture was based
on the scaling relation that may be rewritten as the identity
\qq
\rho_0^{1-\alpha}\,\CQ(\rho_0^{1-\alpha}t;\rho_0,\gamma\rho_0)\vert_{_{D_3}}
\ =\ \CQ(t;1,\gamma)\vert_{_{D_3(\rho_0)}}
\qqq
for $\,D_3(\rho_0)=\rho_0^{1-\alpha-2\beta}D_3\,$ and $\,D_2\,$ unchanged,
see (\ref{resc}). When $\,\rho_0\to\infty\,$ in domain A and B
and $\,\rho_0\to0\,$ in domain C then $\,D_3(\rho_0)\to0\,$ so that we fall
into the situation considered in the scenario (\ref{scen}). The latter
becomes then the assertion that 
\qq
\hbox to 8.8cm{$\rho_0^{1-\alpha}\,\CQ(\rho_0^{1-\alpha}t,\rho_0,\gamma\rho_0)
\,\,1_{\{t<\tau'(\rho_0)\}}$\hfill}& 
\mathop{\longrightarrow}\limits_{\rho_0\to{\infty\atop0}}&
\ \quad\CQ^{fr}(t;1,\gamma)\cr
& &\label{scen'}\\
\hbox to 8.8cm{$D_3^{-1}\rho_0^{2\beta}\,\CQ(D_3^{-1}\rho_0^{2\beta};\rho_0,
\gamma\rho_0)\,\,1_{\{s\geq D_3(\rho_0)\tau'(\rho_0)
\}}$\hfill}&
\mathop{\longrightarrow}\limits_{\rho_0\to{\infty\atop0}}&\ 
\quad\CQ^{\infty}(s;\gamma) 
\nonumber
\qqq
for $\,\tau'(\rho_0)=\CO(|\ln\rho_0|^{\alpha})$. 
\vskip 0.4cm

One of the consequences of such a limiting behavior would be
a bifractal scaling with $\,\rho_0\to\infty\,$ or $\,\rho_0\to0\,$
of the moments of the exit time. Indeed,
\qq
\int\limits_0^\infty t^n\,\,
\CQ(t;\rho_0,\gamma\rho_0)\,dt\ =\ \rho_0^{(1-\alpha)n}\hspace{-0.3cm}
\int\limits_0^{\tau'(\rho_0)}\hspace{-0.15cm} t^n
\,\rho_0^{1-\alpha}\,\CQ(\rho_0^{1-\alpha}t;\rho_0,
\gamma\rho_0)\,dt\,\,\cr
+\ D_3^{-n}\rho_0^{2\beta n}\hspace{-0.6cm}
\int\limits_{D_3(\rho_0)\tau'(\rho_0)}^\infty
\hspace{-0.5cm}s^n\,D_3^{-1}\,\rho_0^{2\beta}\,
\CQ(D_3^{-1}\,\rho_0^{2\beta}s;\rho_0,\gamma\rho_0)\,ds\,.
\qqq
In the regime of extreme values of $\,\rho_0$, \,the first 
term on the right hand side
behaves like $\,\,\rho_0^{(1-\alpha)n}\int\limits_0^\infty t^n\,\CQ^{fr}
(t;1,\gamma)\,dt\,\,$ if we assume (\ref{scen'})
with a sufficiently strong convergence. Similarly the second term 
would behave like
$\,\,D_3^{-n}\rho_0^{2\beta n}\hspace{-0.6cm}\int
\limits_{D_3(\rho_0)\tau'(\rho_0)}^\infty\hspace{-0.5cm}
s^n\,\CQ^{\infty}(s;\gamma)\,ds$. \,If $\,Q^{\infty}(s,\gamma)$ 
is integrable at zero, the first term dominates 
for negative $\,n\,$ and the second one for positive $\,n\,$
when it behaves as $\,\CO(\rho_0^{2\beta n})$. \,The ``minimal'' 
scenario (\ref{scen}) would then imply that
\qq
\langle\, t^n\,\rangle\ =\ 
\cases{\hbox to 3cm{$\CO(\rho_0^{n(1-\alpha)})$\hfill}
{\rm for}\quad n\leq 0\,,\cr\cr
\hbox to 3cm{$\CO(\rho_0^{2\beta n})$\hfill}{\rm for}\quad n\geq 0}
\label{frpr}
\qqq
for large $\,\rho_0\,$ in domain A and B and for small $\,\rho_0\,$ in
domain C.  Such a bifractal behavior is, of course, consistent with
the earlier conjecture (\ref{frtc}). \,The missing probability 
in the frozen case would be given by 
$\,\int\limits_0^\infty\CQ^\infty(s;\gamma)\,ds$.
\,Even in the presence of an intermediate regime in the 
exit time PDF, the scaling (\ref{frpr}) should set in 
for $\,|n|\gg1$. \,The convexity (concavity) of the large (small)
$\,\rho_0\,$ exponent as a function of $\,n$, \,together with its
vanishing at $\,n=0$, \,would then impose the behavior (\ref{frpr})
for all $\,n$.
\vskip 0.4cm

We have tested the scenario (\ref{scen}) numerically. The one-dimensional
velocity field with temporal correlations was obtained
again using the Fourier method. Each Fourier mode $v_k(t)$ was
generated by integrating the corresponding Uhlenbeck-Ornstein
differential equation:
\qq
{dv_k(t)}\ =\ -{v_k(t)\over \tau(k)}dt\,+\,\Big({2E(k)\over
\tau(k)}\Big)^{^{\hspace{-0.08cm}1/2}}dW(t)\,,
\label{UO}
\qqq
where $\,\tau(k)\,$ scales as $\,k^{-2\beta}\,$, the energy spectrum of the
field is denoted by $\,E(k)\propto k^{-1-2\alpha}\,$ and $\,dW(t)\,$ is a
standard Brownian motion.  The stochastic differential equations
(\ref{UO}) were integrated by using a simple Euler scheme of order
${1\over2}$ \cite{KP92}.

\begin{figure}
\begin{center}
\mbox{\hspace{-0.6cm}\epsfig{file=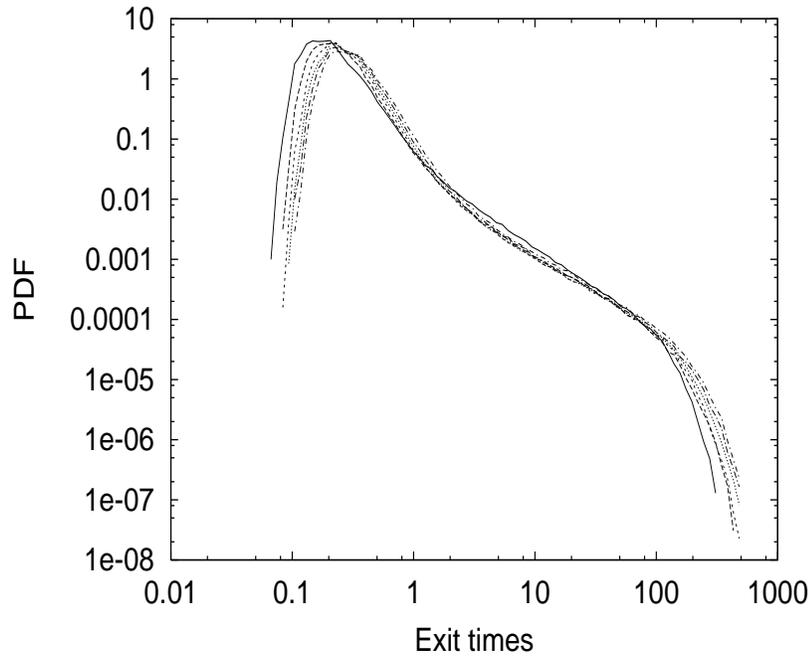,height=9cm,width=11cm}}
\end{center}
\vspace{0.3cm}
\caption{The PDF of the exit times for the flow with $\,\alpha=0.75\,$ and 
$\,\beta=0.3\,$ at six initial separations $\,\rho_0=400, 800, 1200, 1500,
1800, 2100$. \,The resolution is $32768$.}
\label{fig:peter_raw}
\end{figure}

\begin{figure}
\begin{center}
\vskip -0.7cm
\mbox{\hspace{0.4cm}\epsfig{file=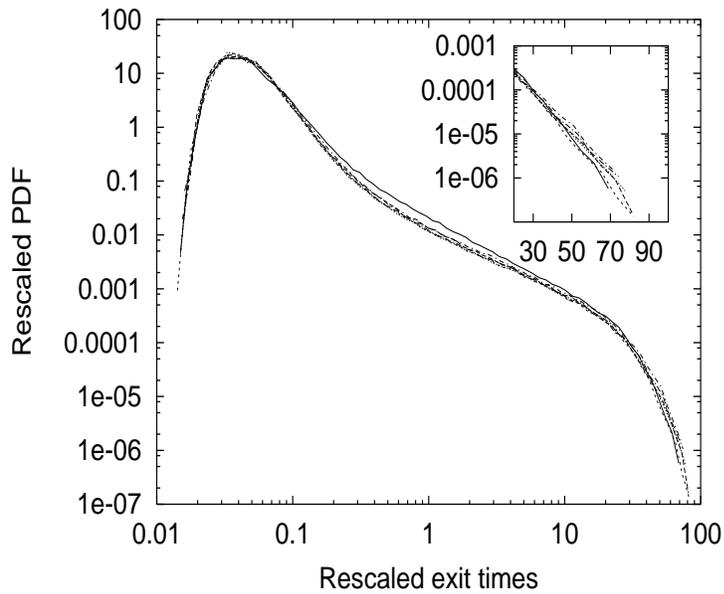,height=9cm,width=11cm}}
\end{center}
\vspace{0.3cm}
\caption{The same PDF as in the previous figure but plotted 
now as $\,\rho_0^{1-\alpha}\CQ\,$ {\it vs} $\,\rho_0^{\alpha-1}t\,$
to display the collapse at small exit times.}
\label{fig:peter_coll_alpha}
\end{figure}

\ 

\vskip -1.5cm

\begin{figure}
\begin{center}
\vskip -0.3cm
\mbox{\hspace{0.2cm}\epsfig{file=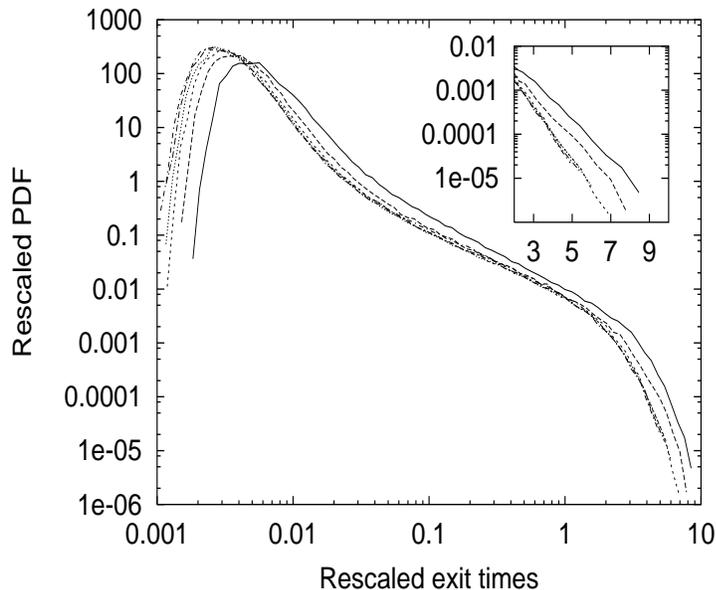,height=9cm,width=11cm}}
\end{center}
\vspace{0.3cm}
\caption{The same PDF as in the previous figure but plotted 
now as $\,\rho_0^{2\beta}\CQ\,$ {\it vs} $\,\rho_0^{-2\beta}t\,$
to display the collapse at large exit times.}
\label{fig:peter_coll_beta}
\end{figure}

\bigskip

\noindent The parameters of the flow where $\,\alpha=0.75\,$
and $\,\beta=0.3$. The exit times for six different initial
separations was measured. Their PDF's $\,\CQ(t;\rho_0,1.05 \rho_0)\,$ are
shown in Fig.~\ref{fig:peter_raw}. According to (\ref{scen'}), by
plotting $\,\rho_0^{1-\alpha}\CQ$ versus $\rho_0^{\alpha-1}t\,$
all the curves should collapse at small exit times, as confirmed in
Fig.~\ref{fig:peter_coll_alpha}. Furthermore, the prediction
(\ref{scen'}) at large exit times is verified in
Fig.~\ref{fig:peter_coll_beta} by collapsing the long-time parts of
the curves by plotting $\,\rho_0^{2\beta}\,\CQ\,$ versus $\,
\rho_0^{-2\beta}t$. \,Note that the curves for the two smallest values of
$\,\rho_0\,$ are not collapsing, in agreement with the previous arguments
predicting that the asymptotic behavior sets in at large $\,\rho_0$'s.
The simulations are consistent with the absence of an intermediate regime
but do not really allow to exclude such a possibility.

\nsection{Eulerian Gaussian velocities: \ sweeping effects}
\label{sec:EUL}

\noindent Let us discuss how the Lagrangian flow changes if the Gaussian
velocity ensemble \ref{vc} is used to model the Eulerian velocities
rather than the quasi-Lagrangian ones. As already mentioned, the main
difference is that, unlike for the Kraichnan model, the separation of
two trajectories is not any more a function of velocity differences
only and it is strongly influenced by large-scale eddies or the so
called sweeping. This effect grows with growing integral scale $\,L\,$
and we shall attempt to study its large $\,L\,$ asymptotics. It seems
to be stronger for small values of $\,\beta$, \,i.e.\,\,for velocities
that are almost frozen at small distances. \vskip 0.5cm

The r.m.s. value of velocity in the ensemble (\ref{vc}) is
proportional to $\,L^\alpha$, \,i.e.\,\,it becomes large for large
$\,L$. \,On the other hand, the r.m.s. equal-time velocity
differences on scales  much smaller than $\,L\,$ are of the order
$\,distance^{\alpha}$. \,In particular, on the scales $\sim
L^\alpha\,$ they are of the order $\,L^{\alpha^2}\ll L^\alpha$. \,
Rewriting the trajectory equation (\ref{Lf}) as\footnote{In
principle, we should add the noise to the trajectory but it does
not play any role on the scales that will be discussed.} \qq \bm
R(t)\ =\ \int\limits_0^t\bm v(s,\bm 0)\,ds\ +\ \int\limits_0^t
[\bm v(s,\bm R(s))-\bm v(s,\bm 0)]\,ds \label{Lfn}\,, \qqq we may
expect that, for fixed $\,t$, \,the first integral is of the order
$\,L^\alpha\,$ and the second of the order $\,L^{\alpha^2}\ll
L^\alpha$. For bounded times, the first integral should then give
the term of the order $\,L^\alpha\,$ of the solution and the second
one, with $\,\bm R(s)\,$ replaced by the approximation $\,\sim
L^\alpha$, \,the term of the order $\,L^{\alpha^2}$. \,More precisely,
let us observe that the Gaussian process with the components
$\,L^{-\alpha}\bm v(t,\bm 0)\,$ and $\,L^{-\alpha^2}
[\bm v(t,L^{\alpha}\bm r)-\bm v(t,\bm 0)]\,$ converges in law 
when $\,L\to\infty\,$ to the
$\,t$-independent Gaussian process $\,(\bm v_0,\,\bm w(\bm r))\,$
with the 2-point functions \qq &&\hbox to 2.8cm{$\langle\,\bm
v_0\,\bm v_0\,\rangle$\hfill}=\ \ D_2\int{1\over
k_1^{d+2\alpha}}\,{d\bm k\over(2\pi)^d}\,,\cr &&\hbox to
2.8cm{$\langle\,\bm w(\bm r)\,\bm w(\bm r')\,\rangle$\hfill}=\ \
D_2\int{(1-\ee^{i\bm k\cdot\bm r})(1-\ee^{-i\bm k\cdot\bm
r'})\over k^{d+2\alpha}}\,{d\bm k\over(2\pi)^d}\,,\cr &&\hbox to
2.8cm{$\langle\,\bm v_0\,\bm w(\bm r)\,\rangle$\hfill}=\ \ 0\,\,.
\label{2pL}\qqq Note the independence of $\,\bm v_0\,$ and $\,\bm
w(\bm r)$. \,It is then natural to conjecture that the following
convergence in law takes place: \qq &&\hbox to 4.6cm{$L^{-\alpha}
\bm R(t)$\hfill}\ \mathop{\longrightarrow}\limits_{L\to\infty}\qquad 
\bm v_0t\,,\label{conj0}\\ &&\hbox to 4.6cm{$L^{-\alpha^2}\Big[\bm
R(t)-\int\limits_0^t\bm v(s,\bm 0)\,ds\Big]$\hfill}\
\mathop{\longrightarrow}\limits_{L\to\infty}\qquad
\int\limits_0^{t}\bm w(\bm v_0s)\,ds\,.\label{conj1}\qqq 
In the frozen case or if $\,\alpha\leq\beta\,$ or 
$\,\alpha>\beta>\alpha(1-\alpha)$, \,the limits describe 
the leading terms in the single trajectory statistics 
for large $\,L$. \,For $\beta\leq\alpha(1-\alpha)$, \,
one should also take into account the term coming from $\,\int\limits_0^t
[\bm v(s,\bm0)-\bm v(0,\bm0)]\,ds\,$ which is of order $\,L^{\alpha-\beta}$.
The dominant term of order $\,L^\alpha\,$ in $\,\bm R(t)\,$ describes the 
ballistic motion with the random velocity of the largest scale eddies that 
sweep the Lagrangian particle along. In Appendix 3 we prove convergence 
(\ref{conj0}) in the frozen one-dimensional model with $\,\alpha={1\over2}$.
\vskip 0.5cm

\begin{figure}
\begin{center}
\vskip 0.3cm
\mbox{\hspace{0.0cm}\epsfig{file=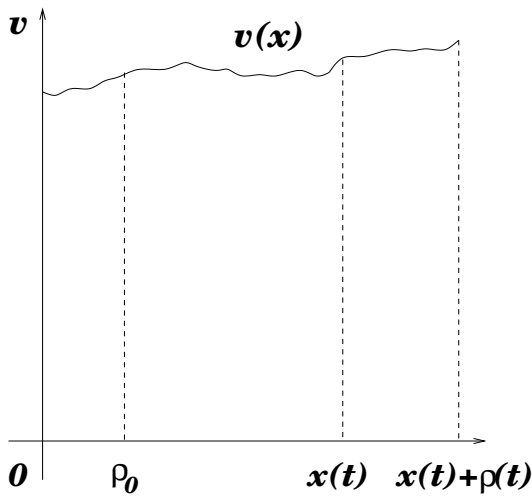,height=6.5cm,width=7cm}}
\end{center}
\vspace{0.3cm}
\caption{Positions of two right-moving particles in $1d$ frozen
velocity}
\label{fig3}
\end{figure}

How does the presence of large scale $\,L\,$ in Eulerian
velocities influence the Lagrangian particle separation? Let us
try to understand this in the one-dimensional frozen model. We
shall consider two particle trajectories $\,x(t)\,$ and
$\,x(t)+\rho_0\,$ \,starting at time zero at zero and
\,$\rho_0>0$, \,respectively, and we shall try to estimate the
behavior of their separation $\,\rho(t)$. \,First notice that
$\,\rho(t)\geq0$, i.e. the order of the particles on the line will
never change. For large $\,L$, \,the dominant events are when the
velocities of the particles and at the points between them are all
of the order $\,L^\alpha\,$ and of the same sign during the time
interval $\,(0,t)$. \,Let us suppose that they are positive, see
Fig.~\ref{fig3} (the case of negative velocities can be treated in a
symmetric way). The crucial fact resulting from the one-dimensional
geometry is the magic identity \qq
\int\limits_0^{\rho_0}{{d\rho}\over{v(\rho)}}\ \ =\
\int\limits_{x(t)}^{x(t)+\rho(t)}{{d\rho}\over{v(\rho)}}\,.
\label{magic} \qqq The left hand side is the time $\,\Delta t\,$
that the first particle takes to reach the initial position of the
second one. The best way to understand the above identity is by
releasing the second particle after the delay $\,\Delta t\,$ so
that both particles move subsequently together. The delay changes
nothing in the movement of the second particle since the velocity
field is frozen. The delayed particle will then be at position
$\,x(t)\,$ at time $\,t\,$ (together with the first particle) and
at position $\,x(t)+\rho(t)\,$ at time $\,t+\Delta t$. But the
right hand side of Eq.\,\,(\ref{magic}) is the time that the
second particle takes to move from $\,x(t)\,$ to $\,x(t)+\rho(t)$.
\vskip 0.5cm

\vskip 0.5cm

\noindent Hence the identity which may be also proven more formally by
noticing that the time derivative of its right hand side vanishes.
Writing for large $\,L$ \qq x(t)\ =\ L^\alpha v_0 t\,+\
\CO(L^{\alpha^2})\,,\qquad v(x(t))\ =\ L^\alpha
v_0\,+\,L^{\alpha^2}w(v_0t)\,+\ \CO(L^{\alpha^3})\,, \qqq see
relations (\ref{conj0}) and (\ref{conj1}), and anticipating that
$\,\rho(t)=\CO(1)$, \,Eq.\,\,(\ref{magic}) may be approximated as
\qq {{\rho_0}\over{L^\alpha v_0}}\ =\ {{\rho(t)}\over{L^\alpha
v_0+ L^{\alpha^2}w(v_0t)}}\,+\,\CO(L^{\alpha^3-2\alpha}) \qqq from
which we infer that \qq \rho(t)-\rho_0\ =\
L^{\alpha^2-\alpha}\rho_0\,v_0^{-1}w(v_0t)\ +\
\CO(L^{\alpha^3-\alpha})\,. \qqq The process $\,w(x)\,$ is the fBm
with the 2-point function (\ref{fbm}) (up to normalization). The
precise conjecture would then assert the convergence in law \qq
L^{\alpha(1-\alpha)}\,[\rho(t)-\rho_0]\ \quad
\mathop{\longrightarrow}\limits_{L\to\infty}\ \quad
\rho_0\,v_0^{\alpha-1}\,w(t)\,. \qqq Note that the above
calculations indicate that not only a single particle motion, but
also the separation of trajectories in the Eulerian frozen
one-dimensional velocity ensemble are dominated by the scale
$\,L\,$ velocities, i.e.\,\,by the large eddy sweeping. The effect 
on the pair dispersion is, however, inverse to that on the single particle
motion. Whereas the latter one becomes very fast for large $\,L$,
\,the trajectory separation becomes essentially frozen to the
initial value in a localization-type effect. It would be
interesting to know if such localizing tendency persists in the
more general Eulerian Gaussian ensembles (\ref{vc}). \vskip 0.5cm

\begin{figure}
\begin{center}
\vspace{-0.9cm}
\mbox{\hspace{-0.5cm}\epsfig{file=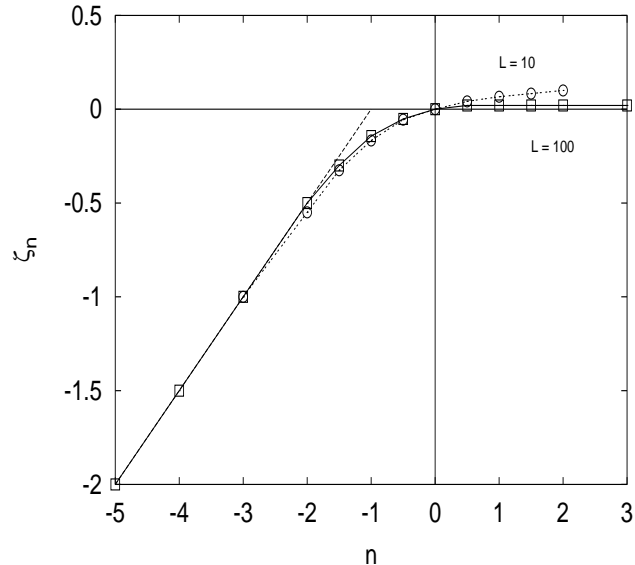,height=7.8cm,width=8.8cm}}
\end{center}
\vspace{0.3cm}
\caption{Scaling exponents of exit-time moments in Eulerian
$1d$ simulations for two different box sizes. Illustration
of the sweeping effects}
\label{fig:euler}
\end{figure}

That the sweeping modifies the pair separation statistics for finite 
$\,L\,$ may be seen in the following way. There is a competition between two 
types of contributions to the dynamics of the pair dispersion $\,\rho(t)$. 
\,The first comes from the configurations where the velocity differences
at distances of order $\,\rho(t)\,$ are much smaller than the 
velocity of each particle. The second one from the opposite regime.
The two contributions may be separated if we fix the initial velocity
$\,v(0)\,$ of the first particle, with $\,v(0)<D_2^{1/2}\rho_0^\alpha\,$
corresponding to the first regime and $\,v(0)>D_2^{1/2}\rho_0^\alpha\,$ to 
the second one. Denote by $\,\CQ(t;\rho_0,\rho_1|\m v(0))\,$ the conditional 
PDF of the exit times for fixed  $\,v(0)$. \,In particular,
$\,\CQ(t;\rho_0,\rho_1|0)\,$ is the quasi-Lagrangian PDF studied
in the $\,L\to\infty\,$ limit in previous Section.
As long as $\rho_0,\,\rho_1\ll L\,$ and $\,v(0)\ll D_2^{1/2}L^\alpha$,
\,the conditional PDF should be approximately $\,L$-independent and,
consequently, it should satisfy the scaling identity
\qq
\mu\,\CQ(\mu t;\mu^\sigma\rho_0,\mu^\sigma\rho_1|\m\mu^{\sigma-1}v(0))
\ \simeq\ \CQ(t;\rho_0,\rho_1|v(0))
\qqq
for $\,\sigma={1\over1-\alpha}$. \,We infer that
\qq
\CQ(t;\rho_0,\rho_1|v(0))\ \simeq\ \rho_0^{\alpha-1}\,\CQ(\rho_0^{\alpha-1}t;
1,\gamma\m|\m\rho_0^{-\alpha}v(0))\,.\label{es1}
\qqq
where, as usually, $\,\gamma={\rho_1\over\rho_0}$. \,Deep in the regime 
$\,v(0)<D_2^{1/2}\rho_0^\alpha\,$ the PDF $\,\CQ(t;\rho_0,\rho_1|v(0))\,$ 
is then essentially quasi-Lagrangian. 
As for the opposite regime, we may use the magic formula
(\ref{magic}) with $\,\rho(t)=\gamma\rho_0$. \,Deep in that regime,
the fluctuations of $\,v(\rho)\,$ in both integrals are small and 
Eq.\,\,(\ref{magic}) reduces to the approximate identity
\qq
v(x(t))\ \simeq\ \gamma\,v(0)
\qqq
from which $\,\rho_0\,$ dropped out and which states that the exit
time $\,t\,$ is the first time when the velocity on the trajectory of
the first particle reaches the value $\,\gamma\,v(0)$.
\,In the scaling regime, we obtain then
\qq
\CQ(t;\rho_0,\rho_1|\m v(0))\ \simeq\ v(0)^{1-1/\alpha}\,
\CQ^{sc}(v(0)^{1-1/\alpha}t;\gamma)\label{es2}
\qqq
which is consistent with (\ref{es1}) in the crossover region $\,v(0)
=\CO(\rho^\alpha)$, \,Even for $\,v(0)\gtrsim D_2^{1/2} L^\alpha\,$
where the scaling breaks, the $\,\rho_0$-independence of 
$\,\CQ(t;\rho_0,\rho_1|\m v(0))\,$ persist so that the contribution
of the region $\,v(0)>D_2^{1/2}\rho_0^\alpha\,$ to the moments of exit time
is approximately $\,\rho_0$-independent for fixed $\,\gamma$.
On the other hand, the contribution of the quasi-Lagrangian regime
$\,v(0)<D_2^{1/2}\rho_0^\alpha\,$ to the $\,n^{\rm th}$-moment is
approximately proportional to
\qq
\int\limits_0^{D_2^{1/2}\rho_0^\alpha}\hspace{-0.2cm}
dv(0)\int t^n\,\CQ(t;\rho_0,\gamma
\rho_0|v(0))\m\,dt\ \simeq\ D_2^{1/2}\rho_0^{\alpha+n(1-\alpha)}
\int t^n\,\CQ(t;1,\gamma\m|\m0)\m\,dt\,.
\qqq
It dominates for small $\,\rho_0\,$ if $\,n<-{\alpha\over1-\alpha}$.
Altogether, we then expect that in the frozen one-dimensional
Eulerian model and for small $\,\rho_0$,
\qq
\langle\,t^n\,1_{\{t<\infty\}}\,\rangle\ =\ 
\CO(\rho_0^{\zeta_n})\quad\ {\rm with\quad\ }
\zeta_n\ =\ \cases{\hbox to 3cm{$\ 
\alpha+n(1-\alpha)$\hfill}{\rm for}\quad\ n
\leq-{\alpha\over1-\alpha}\,,\cr
\hbox to 3cm{$\ 0$\hfill}{\rm for}\quad\ n
\geq-{\alpha\over1-\alpha}\,,}
\qqq
i.e.\,\,again a bifractal situation. The prediction seems to be 
confirmed, at least for large $\,|n|$, \,by numerical simulations, 
see Fig.\,\ref{fig:euler}.
\vskip 1.2cm

\vskip 0.3cm
\nappendix{A}
\vskip 0.3cm

\noindent We shall establish here the estimate (\ref{posa}) on the average
time, given by Eq.\,\,(\ref{avt1+}), that the trajectory takes
to reach the first zero of $\,w(\rho)\,$ between $\,\rho_0$ and
$\,\rho_2$. \,To this end, let us note that for $\,\delta>0$, \qq
\int\limits_{0}^{\infty}\hspace{-0.06cm}\ee^{-{(w+w_2)^2\over2
(\rho_2-\rho)}}{_{\sqrt{2}\,dw_2}\over^{\sqrt{\pi(\rho_2-\rho)}}}\leq
\sqrt{2}\,\ee^{-{w^2\over4(\rho_2-\rho)}}\leq
\sqrt{2}\left({_{\rho_2 }\over^{\rho_2-\rho}}\right)^{\delta}
\ee^{-{w^2\over4(\rho_2-\rho)}}\leq{\sqrt{2}\,(4\delta)^\delta
\ee^{-\delta}}{_{\rho_2^{\,\delta}}\over^{w^{2\delta}}}\,. \qqq
Employing this bound for $\,0<\delta<1/2\,$ and extending the
integral over $\,\rho\,$ in (\ref{avt1+}) to infinity with the
use of the identity \qq
\int\limits_{\rho_0}^\infty\Big(\ee^{-{(w_0-w)^2\over
2(\rho-\rho_0)}}
-\,\ee^{-{(w_0+w)^2\over2(\rho-\rho_0)}}\Big){_{d\rho}\over^{\sqrt{2
\pi(\rho-\rho_0)}}}\ =\ w_0+w-|w_0-w|\,, \label{1la} \qqq we
obtain \qq
\langle\,t_+\,1_{\{w(\rho_0)>0,\,\rho_+\leq\rho_2\}}\rangle\ \leq\
{\sqrt{2}}\,(4\delta)^\delta\,\ee^{-\delta}\,
\rho_2^{\,\delta}\int\limits_0^{\infty}\ee^{-{w_0^2\over2\rho_0}}
{_{dw_0}\over^{\sqrt{2\pi\rho_0}}}\int\limits_0^\infty{_{(w_0+w-|w_0-w|)
\,dw}\over^{w^{1+2\delta}}}\,\,\cr\cr
={_{(2\delta)^\delta\ee^{-\delta}\Gamma(1-\delta)}
\over^{\sqrt{\pi}\,\delta(1-2\delta)}}\,\rho_0^{1/2-\delta}
\rho_2^{\,\delta}\,. \qqq The minimization over $\delta$ gives the
inequality (\ref{posa}).

\nappendix{B}
\vskip 0.3cm

\noindent This appendix is devoted to the spectral analysis of the operator
$\,\CK_-\,$ given by Eq.\,\,(\ref{ck}) with the attractive
potential, pertaining to the long-time behavior of the $\,d=1$,
$\,\alpha={1\over2}\,$ exit times. The two eigen-solutions of
$\,\CK_-\,$ corresponding to an eigenvalue $\,\lambda\,$ may be
expressed by the Whittaker functions
\begin{eqnarray}
\psi_{\lambda}(w)\ =\ M_{{1\over\sqrt{-2\lambda}},\,
{1\over2}}(2\sqrt{-2\lambda}\,\,w)\,,\qquad \varphi_{\lambda}(w)\
=\ W_{{1\over\sqrt{-2\lambda}},\,
{1\over2}}(2\sqrt{-2\lambda}\,\,w)\,.\label{e1}
\end{eqnarray}
The spectrum of $\,\CK_-\,$ on the positive half-line and with the
Dirichlet boundary condition at the origin is composed of the
half-line $\,[0,\infty)\,$ (continuous spectrum) and of discrete
negative eigenvalues $\,E_n=-{1\over2 n^2}\,$ for $\,n=1,2,\dots$
\,The eigenfunctions in the spectrum are $\,\psi_{_E}(w)\,$ where
in (\ref{e1}) for $\,E>0\,$ we choose the square root with
positive imaginary part and for $\,E=E_n\,$ the positive one.
These functions vanish at zero. They are imaginary and oscillating
at infinity for $\,E>0$. \,For $\,E=E_n$, \,they are real 
and decaying exponentially. In the latter case, the two 
eigen-solutions (\ref{e1})
become proportional and may be expressed by the Laguerre
polynomials, similarly as for the three-dimensional Schr\"{od}inger
operator in the attractive Coulomb potential: \qq
&&M_{n,{1\over2}}({_{2}\over^{n}} \,w)\ =\
{_{(-1)^{n-1}}\over^{n!}}\,\,W_{n,{1\over2}}({_{2} \over^{n}}\,w)
\,\,\cr &&=\ {_{2}\over^{n^2}}\,w\, \ee^{-{1\over n}\,
w}\,\,L^1_{n-1}({_{2} \over^{n}}\,w)\ =\
{_1\over^{n!}}\,\,\ee^{z/2}
\,{_{d^{n-1}}\over^{dz^{n-1}}}\left(z^n\,\ee^{-z}\right)\Big|_{z={2
\over n}\,w}\,. \qqq The resolvent kernel of $\,\CK_-\,$ takes the
form \qq (\CK_--\lambda)^{-1}(w_0,w_1)\ =\
{_2\over\CW}\,\cases{\psi_\lambda(w_0)\,
\varphi_\lambda(w_1)\qquad{\rm for}\quad w_0\leq w_1\,,\cr
\varphi_\lambda(w_0)\,\psi_\lambda(w_1)\qquad{\rm for}\quad
w_0\geq w_1\,,} \label{resk} \qqq with the Wronskian \qq \CW\ =\
\varphi_\lambda(w)\,\partial_{_w}\psi_\lambda(w)
-\psi_\lambda(w)\,\partial_{_w}\varphi_\lambda(w)\ =\
{2\sqrt{-2\lambda}/\Gamma(1-{_{1}\over^{\sqrt{-2\lambda}}})}\,,
\label{wr} \qqq where in the expression for the resolvent the
square roots are taken positive for $\,\lambda\,$ sufficiently
negative and continued analytically to the other values of
$\,\lambda\,$ outside the spectrum. The discrete eigenvalues
$\,E_n$ appear as poles in the right hand side of (\ref{resk})
with the residue 
\qq
-\sqrt{-2E_n}\,\psi_{_{E_n}}(w_0)
\,\overline{\psi_{_{E_n}}(w_1)}
\qqq
originating in the zeros of the Wronskian. Along the positive
axis of $\,\lambda$, \,the right hand side of (\ref{resk}) has a
cut 
\qq
{_{\pi i}\over^{E}}\Big(1-\ee^{-{2\pi\over\sqrt{2E}}}\Big)^{-1}\,
\psi_{_{E}}(w_0)\,\overline{\psi_{_{E}}(w_1)}\,.
\qqq
It follows that the spectral density of $\,\CK_-\,$ has the form \qq 
\nu(E)\ =\ \sum\limits_{n=1}^\infty \sqrt{-2E_n}\,\,\delta(E-E_n)
\,+\,{_{1}\over^{2E}}\Big(1- \ee^{-{2\pi\over\sqrt{2E}}}\Big)^{-1}
\qqq 
and that \qq
&&\int\limits_0^\infty\ee^{-|\omega|^2(\rho_1-\rho_0)
\CK_-}(w_0,w_1)\,\,dw_1\ =\ \sum\limits_{n=1}^\infty
{\ee^{\,{|\omega|^2(\rho_1-\rho_0)\over 2\,n^2}}\,{_{1}
\over^{n}}}\,\,M_{n,{1\over2}}({_{2}
\over^{n}}\,w_0)\,\int\limits_0^\infty M_{n,{1\over2}}({_{2}
\over^{n}}\,w_1)\,\,dw_1\,\,\cr &&\qquad-\ \int\limits_0^\infty
dw_1\int\limits_0^\infty {_{\ee^{-|\omega|^2(\rho_1
-\rho_0)E}}\over ^{2E\Big(1-\ee^{-{2\pi\over\sqrt{2E}}}\Big)}}
\,\,M_{{1\over i\sqrt{2E}},\,{1\over2}}(2i\sqrt{2E}\,w_0)\,\,
M_{{1\over i\sqrt{2E}} ,\,{1\over2}}(2i\sqrt{2E}\,w_1)\,\,dE\,.
\qqq Substituting this expression to Eq.\,\,(\ref{eck}), one can
see that the contribution of the ground state of
$\,\CK_-\,$ dominates for $\,\omega=-i|\omega|$ and large
$\,|\omega|\,$ so that \qq \langle\,\ee^{i\omega
t}\,1_{\{t<\infty\}}\,\rangle\,\ =\
{_{2\sqrt{2}}\over^{\sqrt{\pi\rho_0}\,\,|\omega|}}\,\,\ee^{\,{|\omega|^2
(\rho_1-\rho_0)\over2}}\,(1\,+\,\CO(|\omega|^{-2})\,. \qqq
\vskip -0.3cm

\nappendix{C}
\vskip 0.3cm

\noindent As an illustration to Sec.\,\ref{sec:EUL}, we shall prove here the
convergence (\ref{conj0}) for the one-dimensional frozen case of
the Gaussian ensemble (\ref{vc}) of Eulerian velocities with
$\,\alpha={1\over2}$. \,Using the scaling properties of the frozen
velocities, both $\,x(t)\,$ and $\,v_0\,$ may be realized on the
same probability space corresponding to the velocity process
$\,\tilde v(x)\,$ with $\,L\,=1$. This is done by setting \qq x(t)\ =\
L\tilde x(L^{-{_1\over^2}}t)\,,\qquad v_0\ =\ \tilde v(0)\,, \qqq where
$\,\tilde x(t)\,$ is the Lagrangian trajectory in the field
$\,\tilde v(x)\,$ such that $\,\tilde x(0)=0$. \,We shall prove the
(stronger) convergence (\ref{vc}) in the $\,L^2$-norm on the
probability space of $\,\tilde v\,$: \qq
\langle\,[\,L^{_1\over^2}\tilde x(L^{-{_1\over^2}}t)\,-\,\tilde v(0)t\,]^2\,
\rangle\quad \mathop{\longrightarrow}\limits_{L\to\infty}\quad0\,.
\label{exc} \qqq If $\,\tilde v(0)>0\,$ then $\,\tilde x(t)>0\,$ and,
symmetrically, if $\,\tilde v(0)<0\,$ then $\,\tilde x(t)<0$. The
contributions of the two cases to the expectation (\ref{exc}) are
equal so that we may restrict ourselves to the case $\,\tilde v(0)>0$.
It will be more convenient to estimate the expectations of the
exit time $\,\tilde t(x)\,$ of $\,\tilde x\,$ through $\,x>0\,$ related to
$\,\tilde x(t)\,$ by the identity \qq 1_{\{\m\tilde x(t)\geq x\m\}}\ =\
1_{\{\m\tilde t(x)\leq t\m\}}\,. \label{1=1} \qqq Since
$\,\tilde t(x)=\int\limits_0^x{dy\over \tilde v(y)}\,$ if $\,\tilde v>0\,$ on
$\,[0,x)$ and is infinite otherwise, we have easy bounds \qq
1_{\{\m\tilde v_{\rm min}\geq x/t\m\}}\ \leq\ 1_{\{\m\tilde t(x)\leq t\m\}}\ \leq\
1_{\{\m\tilde v_{\rm av}\geq x/t\m\}} \label{1<1<1} \qqq with $\,\tilde v_{\rm
min}\,$ being the minimum of $\,\tilde v\,$ on the interval $\,[0,x]\,$
and $\,\tilde v_{\rm av}={1\over x}\int \limits_0^x\tilde v\,$ its
average value.
Now, with the use of the identity (\ref{1=1}) and integration by
parts, the $\,L^2$-norm on the left hand side of (\ref{exc}) may
be rewritten as \qq 4\int\limits_0^\infty
x\,\,\langle\,1_{\{\,\tilde t(L^{-{_1\over^2}}x)\leq
L^{-{_1\over^2}}t\,\}}\,\rangle\,\,dx\ -\
4t\int\limits_0^\infty\langle\, 1_{\{\,\tilde t(L^{-{_1\over^2}}x)\leq
L^{-{_1\over^2}}t\,\}}\,\tilde v(0)\,\rangle\,\,dx \ +\
t^2\,\langle\,\tilde v(0)^2\,\rangle\,. \label{00} \qqq From the
explicit expressions for the Gaussian field expectations, it
follows that, for $\,\tilde v_{\rm min}\,$ and $\,\tilde v_{\rm av}\,$
standing now for the minimum and the mean of $\,\tilde v\,$ over the
interval $\,[0,L^{-{_1\over^2}}x]$, \qq \lim\limits_{L\to\infty}\
\langle\,1_{\{\,\tilde v_{\rm min}\geq x/t\,\}}\,\rangle \ =\
\lim\limits_{L\to\infty}\ \langle\,1_{\{\,\tilde v_{\rm av}\geq
x/t\,\}}\, \rangle\ =\ \langle\,1_{\{\,\tilde v(0)>x/t\,\}}\,\rangle\,, \qqq
so that also \qq \lim\limits_{L\to\infty}\
\langle\,1_{\{\,\tilde t(L^{-{_1\over^2}}x)\leq L^{-{_1\over^2}}t\,\}}\,
\rangle\ =\
\langle\,1_{\{\,\tilde v(0)>x/t\,\}}\,\rangle \qqq and similarly with the
insertion of $\,\tilde v(0)$. \,It is also easy to show a uniform in
$\,L\,$ bound $\,\, \langle1_{\{\,\tilde v_{\rm min}\geq
x/t\,\}}\rangle\leq\ee^{-C(x/t)^2}$. From the Dominated Convergence
Theorem, the limit of (\ref{00}) is then equal to the expression \qq
4\int\limits_0^\infty x\,\langle\,1_{\{\,\tilde v(0)\geq
x/t\,\}}\,\rangle\,dx\ - \
4t\int\limits_0^\infty\langle\,1_{\{\,\tilde v(0)\geq
x/t\,\}}\,\tilde v(0)\,\rangle\,dx \ +\ t^2\,\langle\,\tilde v(0)^2\,\rangle
 \label{000} \qqq which vanishes in a Gaussian ensemble.
Generalization of this proof to the case
with $\,\alpha\not=0\,$ does not pose much problem.
\eject

\begin{center}
\large\bf References
\vskip -1.7truecm
\end{center}

\end{document}